%% file: ms.tex
\documentclass{aa}
\usepackage{natbib}
\usepackage{graphicx}
\usepackage{txfonts}
\usepackage{xcolor}
\usepackage{hyperref}
\usepackage{url}
\hypersetup{
colorlinks,
 linkcolor={red!80!black},
 citecolor={blue!80!black},
 urlcolor={blue!80!black}
}

\def\nifs{\iso{56}Ni}
\def\cofs{\iso{56}Co}

\def\cm3{cm$^{-3}$}
\def\kms{km~s$^{-1}$}

\def\lsun{L$_{\odot}$}
\def\rsun{R$_{\odot}$}

\def\msun{M$_{\odot}$}

\def\one{\ts {\,\sc i}}
\def\two{\ts {\,\sc ii}}
\def\three{\ts {\,\sc iii}}
\def\four{\ts {\,\sc iv}}
\def\five{\ts {\,\sc v}}

\def\beq{\begin{equation}}
\def\eeq{\end{equation}}

\def\lesssim{\mathrel{\hbox{\rlap{\hbox{\lower4pt\hbox{$\sim$}}}\hbox{$<$}}}}
\def\gtrsim{\mathrel{\hbox{\rlap{\hbox{\lower4pt\hbox{$\sim$}}}\hbox{$>$}}}}

\def\one{{\,\sc i}}
\def\two{{\,\sc ii}}
\def\three{{\,\sc iii}}
\def\four{{\,\sc iv}}
\def\five{{\sc v}}

\def\v1d{{\tt V1D}}
\def\bec{{\tt BEC}}
\def\kepler{{\tt KEPLER}}
\def\mesa{{\tt MESA}}
\def\cmfgen{{\tt CMFGEN}}

\def\ergs{erg\,s$^{-1}$}

\newcommand{\iso}[2]{\ensuremath{^{#1}\rm{#2}}}

\def\aj{AJ}

\def\pasa{PASA}
\def\apj{ApJ}
\def\apjs{ApJS}
\def\apjl{ApJL}
\def\aap{A\&A}
\def\araa{ARA\&A}
\def\aaps{A\&AS}
\def\mnras{MNRAS}
\def\nat{Nature}

\begin{document}

 \title{Supernovae Ib and Ic from the explosion of helium stars}
 \titlerunning{simulations of SNe Ibc}

\author{
Luc Dessart\inst{\ref{inst1}}
\and
  Sung-Chul Yoon\inst{\ref{inst2}}
 \and
 David R. Aguilera-Dena\inst{\ref{inst3},\ref{inst4}}
\and
 Norbert Langer\inst{\ref{inst3},\ref{inst4}}
}

\institute{
Institut d'Astrophysique de Paris, CNRS-Sorbonne Universit\'e, 98 bis boulevard Arago, F-75014 Paris, France.\label{inst1}
\and
    Department of Physics and Astronomy, Seoul National University, Gwanak-ro 1,
    Gwanak-gu, Seoul, 151-742, Republic of Korea.\label{inst2}
\and
Argelander-Institut f\"{u}r Astronomie, Universit{\"a}t Bonn, Auf dem H\"{u}gel 71, 53121, Bonn, Germany.\label{inst3}
\and
Max Planck Institut f{\"u}r Radioastronomie, Auf dem H{\"u}gel 69, 53121 Bonn, Germany.\label{inst4}
  }

   \date{Received; accepted}

  \abstract{
Much difficulty has so far prevented the emergence of a consistent scenario for the origin of Type Ib and Ic supernovae (SNe). Either the SN rates, or the ejecta masses and composition were in tension with inferred properties from observations. Here, we follow a heuristic approach by examining the fate of helium stars in the mass range 4 to 12\,\msun, which presumably form in interacting binaries. The helium stars are evolved using stellar wind mass loss rates that agree with observations, and which reproduce the observed luminosity range of galactic WR stars, leading to stellar masses at core collapse in the range 3 to 5.5\,\msun. We then explode these models adopting an explosion energy proportional to the ejecta mass, roughly consistent with theoretical predictions. We impose a fixed \nifs\ mass and strong mixing. The SN radiation from 3 to 100 d is computed self-consistently starting from the input stellar models using the time-dependent non-local thermodynamic equilibrium radiative-transfer code \cmfgen. By design, our fiducial models yield very similar light curves, with a rise time of about 20 d and a peak luminosity of $\sim$\,10$^{42.2}$\,\ergs, in line with representative SNe Ibc. The less massive progenitors retain a He-rich envelope and reproduce the color, line widths, and line strengths of a representative sample of SNe Ib, while stellar winds remove most of the helium in the more massive progenitors, whose spectra match typical SNe Ic in detail. The transition between the predicted Ib-like and Ic-like spectra is continuous, but it is sharp, such that the resulting models essentially form a dichotomy. Further models computed with varying explosion energy, \nifs\ mass, and long-term power injection from the remnant show that a moderate variation of these parameters can reproduce much of the diversity of SNe Ibc. We conclude that stars stripped by a binary companion can account for the vast majority of ordinary Type\,Ib and\,Ic SNe, and that stellar wind mass loss is the key to remove the helium envelope in the progenitors of SNe\,Ic.
}

\keywords{
  radiative transfer --
  radiation hydrodynamics --
  supernovae: general
}
   \maketitle

\section{Introduction}

Binary-star evolution has been recognized as an essential ingredient for the production of Type I core-collapse supernovae (SNe) since the 80's \citep{wheeler_levreault_85,wheeler_ibc_87,ensman_woosley_88,podsiadlowski_92,wlw_mdot_95,vanbeveren_bin_98,wellstein_langer_99,dessart_11_wr,smith_11_sn_stat,langer_araa_mdot_12}.  This result was required to explain the frequencies of SNe Ibc (as well as SNe IIb), their  short rise times to maximum light of about 20\,d, and the narrow light-curve widths around bolometric maximum (see, e.g., the analyses of large samples by \citealt{drout_11_ibc}, \citealt{taddia_ibc_15}; \citealt{lyman_ibc_16}, or \citealt{prentice_ibc_16,prentice_snibc_19}). Such features are best explained by invoking low mass ejecta, stemming from a low or moderate mass He star. An obvious progenitor of this scenario is therefore a $10-25$\,\msun\ star in a binary system, which lost its H-rich envelope through interaction with a companion  \citep{eldridge_08_bin,yoon_ibc_10}.

The He-star mass correlates with its luminosity, which controls the strength of the radiatively-driven wind at the origin of the subsequent wind mass loss \citep{langer_wr_mdot_89,grafener_mdot_11}. For low He-core masses, the wind mass loss rate is weak and may not cause much mass loss at all.  Being of low mass and rich in He, the progenitor star may explode as a SN Ib, or a SN IIb if some residual H is present at the surface. Depending on the mass loss rate, the winds of higher mass He-stars may or may not peel off the He-rich outer layers and produce a SN Ic (assuming that He deficiency is a prerequisite for a Type Ic classification; \citealt{liu_snibc_15}).

For example, the mass loss rates used in the study of \citet{yoon_ibc_10}, which explored the stripped component of detailed binary evolution models, were too weak to produce the He-poor progenitors required for a Type Ic classification (these mass loss rate calibrations were taken from \citet{hamann_mdot_82} and \citet{hamann_wr_95}, but scaled down by a factor 5 or 10). Indeed, out of 45 simulations, only two of their models are likely to make a Type Ic SN. Their lowest mass model produces a star at death of only 1.6\,\msun, which would produce an ultra-stripped SN Ic \citep{tauris_ulstr_13}. Their other models,  for  $12-25$\,\msun\ binary stars, which are favored by the initial mass function and also suitable to produce low-mass He stars,  would produce SNe\,Ib (or SNe\,IIb).

In the study of \citet{yoon_ibc_10}, not all binary parameters were explored. For example, the so called Case BB mass transfer after core He exhaustion is an efficient means to peel off the He-rich layers of a He star  \citep{dewi_bin_02,dewi_pols_03}. Such a Case BB mass transfer has been invoked to explain the Type Ic SN\,1994I and its progenitor \citep{nomoto_94I_94}. Although SN\,1994I is considered a standard SN Ic, even a prototype \citep{modjaz_etal_14}, the required binary configuration is probably too rare to produce the majority of SNe Ic.

Recently, \citet{yoon_wr_17} studied the evolution of He stars with mass loss. These He stars are expected to arise primarily from interacting binaries. \citet{yoon_wr_17} argued that the Wolf-Rayet (WR) mass loss rates that influence the subsequent evolution may have been underestimated. Indeed, higher WC mass loss rates appear essential to explain the sizable population of relatively faint WC stars \citep{sander_wc_12}. The mass-luminosity relation during He-core burning \citep{woosley_he_19} implies that these fainter WC stars must be of a lower mass than previously predicted. A reduction in the final mass of WC stars (at least for a subset of these stars) could help resolve the mass discrepancy that currently holds between the high WC star masses at death (see discussion and references in \citealt{yoon_wr_17}) and the relatively low SNe Ic ejecta masses inferred from observations.

Much debate has taken place over the last decades on the possibility of hiding He, and in particular below what threshold He mass He\one\ lines would be invisible \citep{swartz_ib_93,d12_snibc,hachinger_13_he}. Unfortunately, the detection of He\one\ lines in SN spectra is often hard to secure, and this is not limited to SNe Ibc since the problem also affects Type II SNe. Typical SN II-P show at best a weak He\one\,5876\,\AA\ line at early times, but as soon as the recombination phase starts, He\one\ lines are absent in the optical, even though He makes about 35\% of the H-rich envelope mass. Evidently, several solar masses of He go un-noticed in garden-variety SNe II-P. In these SNe II-P, it is the presence of H (essentially at a level set by Big-Bang nucleosynthesis) that implies indirectly the presence of He.

In stripped-envelope core-collapse SNe, the situation is different. In rare instances,  SNe IIb or Ib show a similar blue optical color at early times to what is observed in SNe II-P (e.g. SNe\,1993J or 2008D; \citealt{matheson_93j_00a}; \citealt{soderberg_08D}; \citealt{modjaz_08D}), indicative of a large gas temperature suitable for the production of He\one\ lines through photoionization and recombination. Unfortunately, these blue optical spectra are also quasi-featureless so that He\one\ lines are not evidently present. Instead, He\one\ lines are seen at the recombination epoch, when the SN radiation peaks in the optical and the SN photospheric temperature is around 7000\,K. Under such conditions, the production of He\one\ lines requires non-thermal excitation \citep{lucy_91,swartz_ib_91}. This process is controlled by the amount of \nifs\ and the mixing of \nifs\ in velocity space \citep{d12_snibc}. The mass fraction of He is also a fundamental quantity since the decay power (and the associated non-thermal rates) scales with the species' abundance. So, if a given mixture contains 10\% of He, only 10\% of the decay power will go to He atoms. Hence, not only the low abundance of He disfavors the production of strong He\one\ lines, but the sharing of the decay power between species also means that non-thermal rates affecting He will also be low. In contrast, if the He mass fraction is 90\,\% or more, not only the large He abundance fosters the production of strong lines, but He also absorbs nearly the entire decay power, favoring the non-thermal excitation of He\,\one\ lines \citep{d12_snibc}.

Observations present further biases and challenges. Line overlap (for example of Na\one\,D with He\one\,5876\,\AA) implies that most observed features are blends of multiple lines. Weak lines are also hard to detect unless the signal-to-noise ratio is high. Fast expansion and the associated Doppler broadening yields weaker lines more likely affected by overlap with lines from other atoms or ions. Despite all these difficulties, numerous SNe exhibit strong He\one\ lines. Notorious He\one\ lines in the optical are at 5876, 6678, and 7065\,\AA\ (for example in the CfA sample, SNe 2004gq, 2004gv, or 2005hg;  \citealt{modjaz_etal_14}), but other He\one\ lines in the blue part of the optical are also predicted and observed (see \citealt{dessart_11_wr} and below).  When an ambiguity on He detection arises, it concerns objects in which the He\one\ lines are weak, and sometimes vanishingly so. Hence, there is a clear dichotomy amongst Type I SNe from massive star explosions, with events showing strong He\one\ lines and all the rest.

Massive star evolution predicts the presence of He even in the most evolved WR stars (see, for example, \citealt{meynet_gridevol_94}), and the observations of these stars confirms this (see, for example, \citealt{kingsburgh_wo_95}; \citealt{crowther_wo_02}). But the separation of observed WR stars between WN stars (atmospheric composition inferred to be rich in He and N, with a He mass fraction of 90\,\% or more; \citealt{crowther_wne_95,crowther_wnl_95}) and the WC stars (atmospheric composition inferred to be rich in C and O, with a He mass fraction at the 10\,\% level; \citealt{koesterke_wc_95}; \citealt{grafener_wc_98}; \citealt{hm99}; \citealt{crowther_wc_02}; \citealt{sander_wc_12})  offers a possible explanation for a strict dichotomy between SNe Ib and Ic. The surface composition of observed WC stars also suggest that the absence of He is not expected, and perhaps not required (to produce a SN Ic), in any massive star at core collapse.

 In this work, we focus on ``standard" SNe Ibc, or what progenitors and explosions produce the majority of SNe Ibc. We present radiative transfer simulations of He-star explosions that, for the first time, match the properties of the majority of SNe Ibc, reproducing simultaneously their light curves out to $\sim$\,100\,d and their spectra around maximum light. Rather than fitting one specific event, we aim to produce models that overlap with the parameter space occupied by such ``standard" SNe Ibc and that reproduce the observed dichotomy between the SN classification as Type Ib or Ic. These SN simulations are based on a physically consistent set of stellar-evolution calculations for He stars, made to approximately mimic the evolution of previously H-rich stars that most likely lost their H-rich envelope by binary mass exchange. Unlike all previous studies, we address the suitability of the models both against the photometric properties and the spectroscopic properties using non-local thermodynamic equilibrium (non-LTE) time-dependent radiative transfer calculations that encompass the ejecta evolution from 3 to 50--100\,d after explosion. \citet{woosley_he_19} and \citet{ertl_ibc_20} performed similar simulations for the evolution and explosion of He stars (with a number of differences in the approach) but these were limited to the computation of SN bolometric light curves and a coarser comparison to observations than presented here.

In the next section, we present the numerical approach for computing our stellar evolution models, the explosion models, and how we follow the evolution of the SN radiation with non-LTE time-dependent radiative transfer. We then describe the light curve (section~\ref{sect_lc}) and the spectral (section~\ref{sect_spec}) properties of our ejecta models. In section~\ref{sect_comp_obs}, we confront our model results to a selection of observed SNe Ib and Ic. In section~\ref{sect_disc} we discuss the origin of some degeneracies in the modeling of SNe Ibc. In section~\ref{sect_conc}, we present our conclusions and discuss the implications of our results on the understanding of SNe Ib and Ic.

\begin{figure}
\includegraphics[width=\hsize]{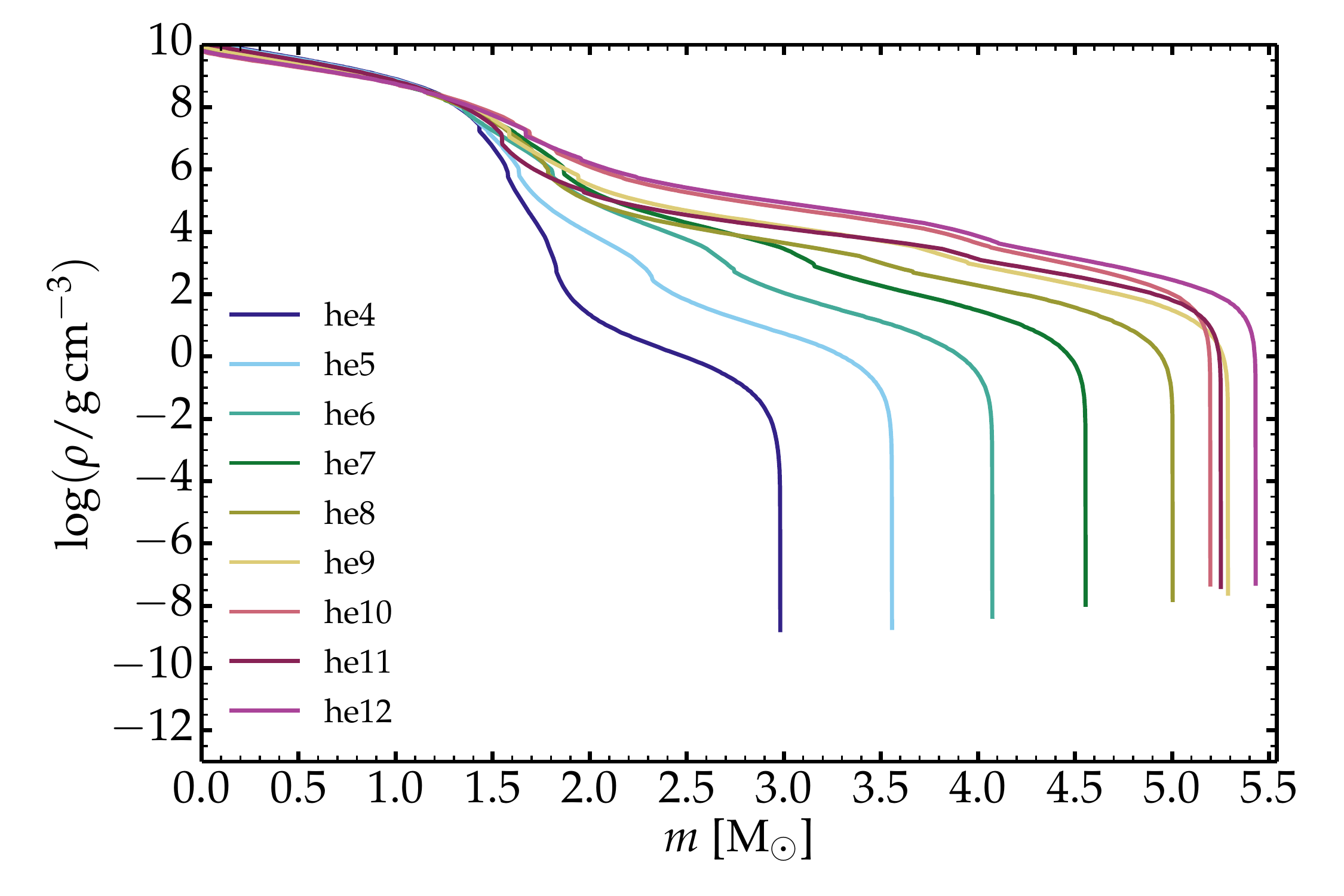}
\includegraphics[width=\hsize]{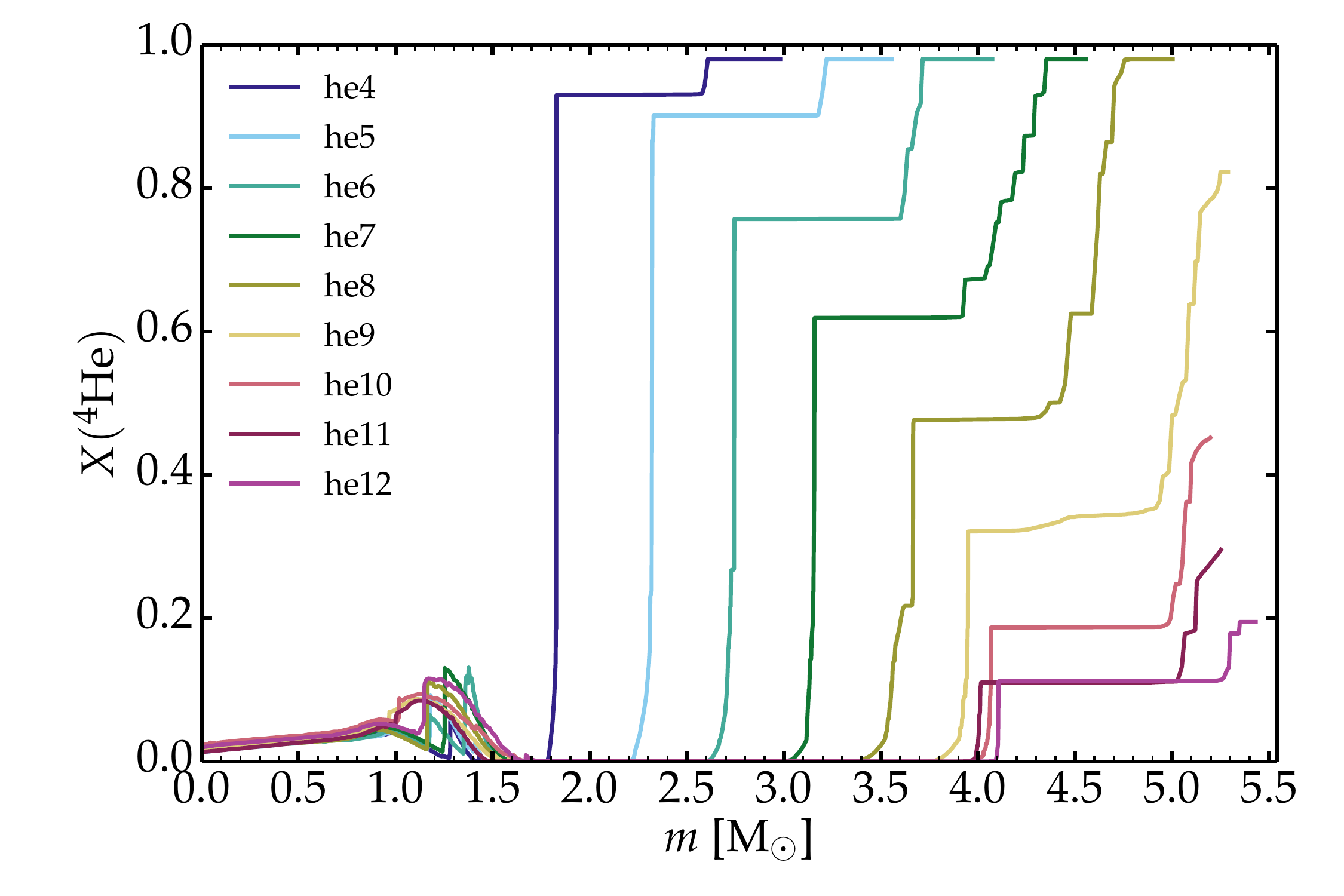}
\includegraphics[width=\hsize]{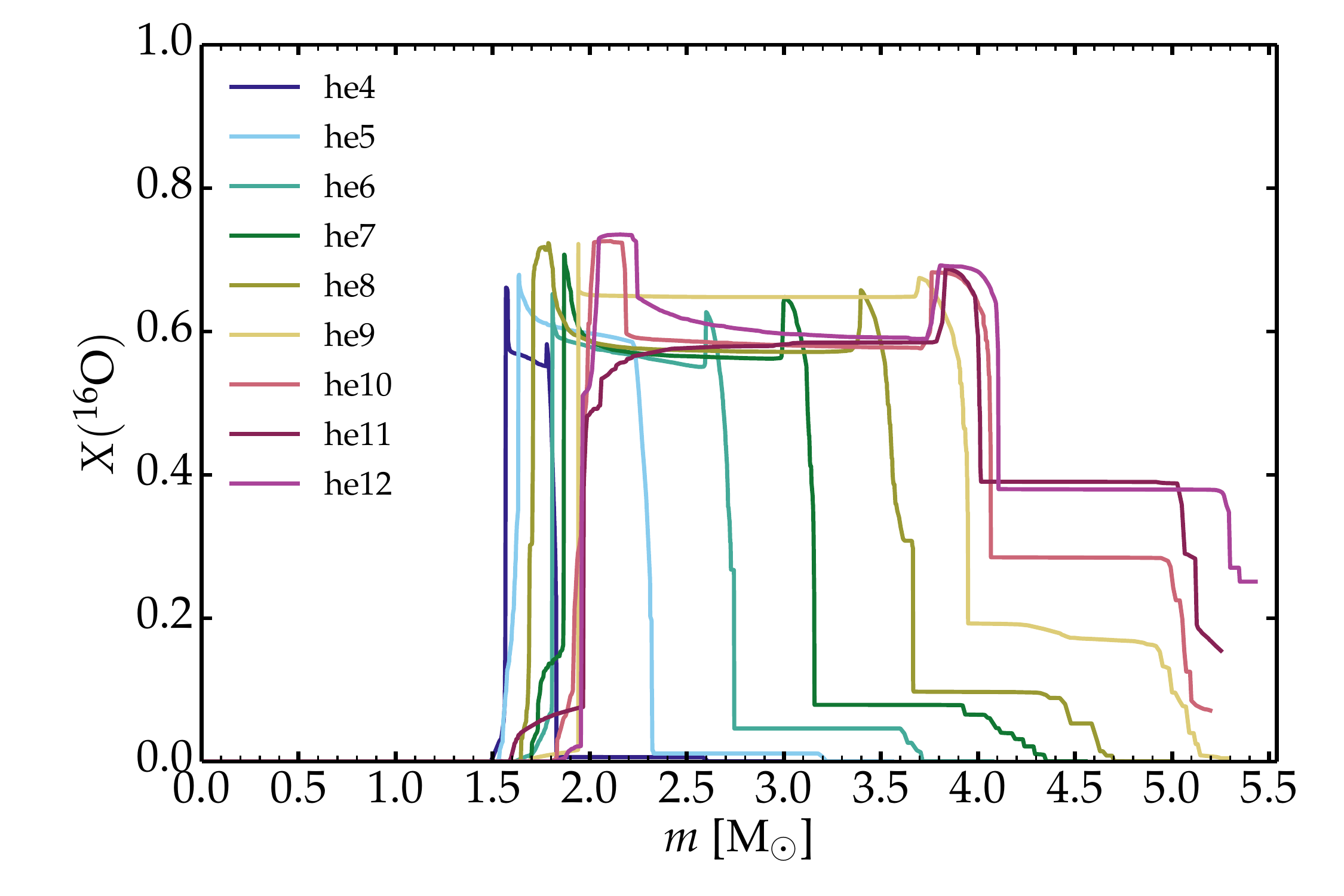}
\caption{Properties of our set of He-star progenitor models he4\,$-$\,he12 at the onset of core collapse. The top panel shows the density structure, followed by the He and O mass fraction, all shown versus Lagrangian mass.
\label{fig_prog}
}
\end{figure}

\section{Numerical approach}
\label{sect_num}

\subsection{Stellar evolution until core collapse with \mesa}
\label{sect_prog}

Our calculations start with pure He-star models with initial masses in the range 4 to 12\,\msun, which are evolved in time using the \mesa\ code \citep[version 11554]{mesa1,mesa2,mesa3,mesa4}. We implicitly assume that these stars stem from 14\,$\cdots$\,32\,\msun\ main sequence stars, and that their hydrogen envelopes have been completely removed by a binary companion once core He burning begins. Detailed binary evolution models show that mass transfer due to Roche-lobe overflow does remove most but not all of the hydrogen envelope \citep{gilkis_mdot_IIb_19}. However, for the metallicity and mass range considered here, assuming that the ensuing strong stellar wind blows off any remaining hydrogen within a short time remains a good approximation \citep{laplace_he_20}. Notably, we do not consider the lowest mass helium stars that could still produce SNe despite their low mass, and that may \citep{yoon_ibc_10} or may not \citep{antoniadis_ia_20} retain their hydrogen.

Whereas for the lower mass He stars considered here, a single star origin appears unlikely, it cannot be excluded that single stars with initial masses near 30\,\msun\ may lose their envelope quickly after core hydrogen exhaustion \citep{smith_owocki_06,petrov_mdot_16}. Corresponding single stars would produce the same SNe as we predict from our He star models, and merely augment their number. Possibly, some single stars in the upper part of the mass range  considered here evolve into red supergiants, where stellar wind mass loss may remove most of the H-rich envelope during core helium burning. The winds during the ensuing WR stage may or may not remove the remaining hydrogen and helium layers. This would lead to larger final masses \citep{woosley_he_19}, some of which may form black holes \citep{langer_bh_20} or Type\,IIb SNe \citep{claeys_IIb_11}. While these objects may contribute to the diversity of stripped envelope SNe, in view of the prevalence of binarity for massive stars \citep{sana_bin_12} we consider binary evolution to be the favored scenario for the formation of our initial He star models.

The initial chemical composition of our models is set to mimic the He-core composition at the end of the main sequence of a 25\,\msun\ star: $X(\mathrm{^4He}) = 0.98$, $X(\mathrm{^{12}C}) = 3.40\times10^{-4}$, $X(\mathrm{^{14}N}) = 1.30\times10^{-2}$, $X(\mathrm{^{16}O}) = 3.4\times10^{-4}$. Elements that are not affected by H burning have the standard values of \citet{gs98} at a metallicity $Z=0.02$. The CNO equilibrium abundances would somewhat differ for different temperatures and hence for different He-star masses, but this detail is not important for our purpose. We adopt the Schwartzschild criterion for convection with a mixing-length parameter of two. We assume no overshooting. Nuclear burning is treated with the network {\tt approx21.net}. All simulations are evolved until the maximum infall velocity of the iron core reaches 1000\,\kms.

For the mass loss rate of such He stars, we follow the prescription by \citet{yoon_wr_17}, which distinguishes mass loss rates from WN and WC/WO type stars, and has a simple interpolation between the two. The adopted WN mass loss rate is based on the study of LMC WN stars by \citet{hainich_wn_14} and includes a scaling by  $Z_\odot^{0.5}$. The WC and WO mass loss rates are based on the study of \citet{tramper_wc_16}. This mass loss rate is significantly higher than the WN mass loss rate for a given luminosity, which tends to exacerbate the composition contrast between WN and WC stars at the pre-SN stage. The stellar wind parameter $f_\mathrm{w}$ is set to 1.6, which corresponds to a wind clumping factor of four ($f_\mathrm{w} = 1.0$ corresponds to a wind clumping factor of 10). We find that $f_\mathrm{w} = 1.6$ can reproduce the luminosity distribution of WN and WC stars in our Galaxy and the LMC \citep{yoon_wr_17}.

Our simulations were performed for He stars with an initial mass covering from 4 to 12\,\msun\ in 1\,\msun\ increments. The model nomenclature is such that he$M$ refers to the He star model with an initial mass of $M$\,\msun. Table~\ref{tab_prog} gives a summary of properties for each of these models at the onset of core collapse. Figure~\ref{fig_prog} illustrates the density stratification, and the composition profile for He and O versus Lagrangian mass. In this model sequence, the surface helium abundance and the He-rich shell mass decrease in favor of a growing surface C/O abundance and CO core mass. The final star mass grows from 2.98 (he4 model) up to 5.43\,\msun\ (he12 model), thus producing lower final masses than previously obtained with alternate mass-loss prescriptions. As discussed by \citet{yoon_wr_17}, these models also produce faint WC stars at death, with $\log L_\star$ in the range $4.6-5.2$. The final mass and luminosity of these models appear similar to those of \citet{ertl_ibc_20} for their ``1.5 $\times$ $\dot{M}$'' prescription (see their Table~1).

Our models are evolved at solar metallicity. Metallicity might play a role, though probably secondary, in the evolution of binaries, and thus in the formation of stripped-envelope stars \citep{gilkis_mdot_IIb_19,laplace_he_20,shenar_wr_20}. However, metallicity will probably influence primarily the WR mass loss rate so that higher metallicities will favor the production of SNe Ic relative to Ib \citep{georgy_snibc_09}. Although there is no strong metallicity dependence affecting the number ratio of SNe Ibc versus SNe II, SNe Ic seem to be more prevalent than SNe Ib at higher metallicity \citep{modjaz_ccsn_z_11,anderson_env_ccsn_15}.

\input{table_prop_mesa_heprog.tex}

\input{table_prop_ejecta.tex}

\subsection{Radiation-hydrodynamics simulations of the explosion with \v1d}
\label{sect_hydro}

The progenitor models described in the previous section are used as initial conditions in the radiation hydrodynamics code \v1d\ \citep{livne_93,dlw10a,dlw10b}. Unlike \citet{ertl_ibc_20}, we do not follow self-consistently the neutrino-driven explosion so that our explosion energies (as well as power versus time or mass cut) are prescribed rather than obtained from first principles (the explosion is still designed at some level in the 1D simulations of \citealt{ertl_ibc_20}). Immediately above the outer edge of the iron core (which increases from about 1.5 to 1.8\,\msun\ from model he4 to model he12), we initiate a thermal bomb by depositing energy in the innermost 0.05\,\msun\ for a duration of 0.2\,s. The total energy deposited that we prescribe corresponds to the envelope binding energy plus an energy per unit ejecta mass of $ 5 \times 10^{50}$\,erg\, $M_\odot^{-1}$. With this scaling, we aim to produce ejecta for models he4 to he12 that have the same $E_{\rm kin} / M_{\rm ej}$. Our values of $E_{\rm kin}$ are representative of those obtained in the more realistic simulations of \citet{ertl_ibc_20} and typical of what may obtain in core collapse SNe like 1987A \citep{sn1987A_rev_90}.

Explosive nucleosynthesis is very sensitive to the position of the mass cut, the core density structure, the power and the total energy of the explosion so the \nifs\ mass differs between  our \v1d\ simulations. Since our explosion setup is artificial, we take the freedom to reset the \nifs\ mass to be 0.08\,\msun\ in all models at 10\,s after the explosion trigger. Having a fixed \nifs\ mass in this he4\,$-$\,he12 sequence facilitates the comparison between models. It is motivated also by the desire to reproduce the properties of garden-variety SNe Ibc, whose light curves are often undistinguishable \citep{drout_11_ibc}. For simplicity, we keep the same nomenclature for these nine ejecta models as for the nine progenitor models, i.e. we stick to he4 to he12 (variants of this reference set of explosion models will be given the same root name with an additional suffix).

The ejecta mass $M_{\rm ej}$ (which is just $M_{\rm f} - M_{\rm Fe\,core}$; our models experience no fallback) increases monotonically in our he4\,$-$\,he12 model sequence from 1.5 to 3.6\,\msun. Since the same power per unit mass is used in our approach, the ejecta kinetic energy increases from 0.75 to 2.1$\times$\,10$^{51}$\,erg in the he4\,$-$\,he12 model set (variations occur because of the nuclear energy released by explosive burning, which is greater in higher mass progenitors). The mean ejecta velocity $V_m = \sqrt{2 E_{\rm kin}/M_{\rm ej}}$ is about 7200\,\kms\ for all models. Our ansatz is that higher mass progenitors explode with a higher energy and yield a higher ejecta kinetic energy at infinity (despite the growing binding energy of their core). This correlation is supported by the simulations of \citet{ertl_ibc_20} but only up to He core masses of about 6\,\msun. We assume that it holds all the way to our most massive progenitor model he12.

In all simulations, some \nifs\ mixing is enforced by using a boxcar with a width set to $M_{\rm ej}/3$. Since the model sequence he4\,$-$\,he12 is also characterized by the same $E_{\rm kin} / M_{\rm ej}$, models he4 to he12 have the same \nifs\ profile in velocity space, modulo an offset in magnitude due to the difference in $M_{\rm ej}$. In other words, $X(^{56}{\rm Ni}) \times M_{\rm ej}$ is the same for models he4 to he12. This mixing is therefore strong in all our models, bringing \nifs\ all the way to the outermost ejecta layers. The short rise times and large peak luminosities of SNe Ibc suggest that weak mixing is unlikely in SNe Ibc. Simulations for varying levels of mixing have been presented earlier (\citealt{d12_snibc,D15_SNIbc_I}; see also \citealt{yoon_ibc_color_19}).

As discussed in a number of recent papers, explaining the SNe Ibc brightness during the photospheric phase or early nebular phase exclusively by \nifs\ decay power requires a \nifs\ mass that is on average larger than for SNe II \citep{drout_11_ibc,lyman_ibc_16,prentice_ibc_16,prentice_snibc_19,meza_anderson_ni56_20,sharon_kushnir_20}. In the simulations of \citet{ertl_ibc_20}, the \nifs\ mass is typically below 0.1\,\msun\ and thus in tension with the value inferred for about half of SN Ibc observations. Part of this tension arises from a few outliers with erroneous estimates of the \nifs\ mass, for example because of an overestimate of the reddening (e.g., SN\,2005hg claimed by \citealt{lyman_ibc_16} to have 0.66\,\msun\ of \nifs\ but this estimate relies on an adopted reddening $E(B-V)$ of 0.685\,mag, when in fact the reddening is low and probably about 0.1\,mag -- see below). However, even without these outliers, an offset remains in the distribution of \nifs\ masses between SNe Ibc and SNe II-P \citep{anderson_nifs_19,meza_anderson_ni56_20}.

Although the main focus of the paper was on reproducing the properties of the majority of SNe Ibc at the lower end of the peak-luminosity distribution, we also investigated possible causes for more luminous SNe Ibc. Indeed, the neutrino mechanism of core-collapse explosion of He stars yields SN ejecta that asymptote to a peak luminosity of about $10^{42.2}$\,\ergs\ \citep{ertl_ibc_20}, while ``standard" SNe Ibc are observed with inferred peak luminosities up to about $10^{42.6}$\,\ergs. For these more luminous SNe Ibc, we tried various options including increasing the ejecta kinetic energy only, increasing both the ejecta kinetic energy and the \nifs\ mass, and finally injecting power from the compact remnant. This power may originate from fallback accretion (see for example \citealt{dexter_kasen_13}) or from dipole-radiation by a magnetized and rotating neutron star (see for example \citealt{KB10}) -- see also discussion in \citet{ertl_ibc_20}. Both processes have a similar time dependence, in particular at late times of weeks to months (as $1/t^{5/3}$ for fallback compared to $1/t^2$ for dipole radiation from a magnetar). For the same energy input, both processes may thus yield a similar impact on the SN ejecta and its radiation. For practical reasons (magnetar power is implemented in \cmfgen; \citealt{d18_iptf14hls}), we parametrize some simulations with a power from a slowly rotating magnetar.

For these variants, we thus considered the following additional models. In models he4ep (he12ep), the total energy deposited and the enforced \nifs\ mass are twice that in model he4 (he12). In models he4epl (he12epl), the total energy deposited is twice that in model he4 (he12), but the \nifs\ is left at 0.08\,\msun. For models he4, he9, he12, two additional variants were done for each by including magnetar power with  $B_{\rm pm} = 10^{15}$\,G and either $E_{\rm pm} = 10^{49}$\,erg  (model with suffix pml) or $E_{\rm pm} = 2 \times 10^{49}$\,erg  (model with suffix pmu). The magnetar with the characteristics ``pml" given above was also included in model he12eppml, which is otherwise the same as model he12ep.

\subsection{Time dependent non-local thermodynamic equilibrium radiative transfer with \cmfgen}
\label{sect_rt}

Using the set of ejecta models produced with \v1d\ and described in the previous section, we solve the non-LTE radiative transfer problem with \cmfgen\ \citep{hm98,DH05a,HD12}. We start the \cmfgen\ simulations at about 3.5\,d after explosion since we wish to focus on the epochs around maximum. Most simulations are stopped at 70\,d after explosion, while a few are continued until 100\,$-$\,150\,d to track the late-time brightness evolution indicative of the magnitude of $\gamma$-ray escape. Metals not treated in the network {\tt approx21.net} during the pre-SN evolution are given an abundance at the solar metallicity value.

 We treat non-thermal processes as per normal \citep{d12_snibc,li_etal_12_nonte}. We limit the radioactive decay to the \nifs\ chain. For simplicity, we compute the non-local energy deposition by solving the radiative transfer equation with a grey absorption-only opacity to $\gamma$-rays set to 0.06 $Y_{\rm e}$\,cm$^2$\,g$^{-1}$, where $Y_{\rm e}$ is the electron fraction.

 The model atom includes He\one\ (40,51), He\two\ (13,30), C\one\ (14,26), C\two\ (14,26), N\one\ (44,104), N\two\ (23,41), O\one\ (19,51), O\two\ (30,111), Ne\one\ (70,139), Ne\two\ (22,91), Na\one\ (22,71),  Mg\one\ (39,122), Mg\two\ (22,65), Al\two\ (26,44), Al\three\ (17,45), Sc\one\ (26,72), Sc\two\ (38,85), Sc\three\ (25,45), Si\one\ (100,187), Si\two\ (31,59), Si\three\ (33,61),
 S\one\ (106,322), S\two\ (56,324), S\three\ (48,98), Ar\one\ (56,110), Ar\two\ (134,415), K\one\ (25,44),  Ca\one\ (76,98), Ca\two\ (21,77), Ti\two\ (37,152), Ti\three\ (33,206), Cr\two\ (28,196), Cr\three\ (30,145), Cr\four\ (29,234),  Fe\one\ (44,136), Fe\two\ (275, 827), Fe\three\ (83, 698), Fe\four\ (51,294), Fe\,\five\ (47,191), Co\two\ (44,162), Co\three\ (33,220), Co\four\ (37,314), Co\,\five\ (32,387),  Ni\two\ (27,177), Ni\three\ (20,107), Ni\four\ (36,200), and Ni\,\five\ (46,183). The numbers in parenthesis correspond to the number of super levels and full levels employed (for details on the treatment of super levels, see \citealt{hm98}).

When comparing to observations, we consider the quasi-bolometric light curves out to $50-100$\,d after explosion but we analyze the spectral properties at one epoch, early after bolometric maximum. Spectra at that time are usually available in SNe Ibc (see, e.g., \citealt{modjaz_etal_14}). Starting at or early after maximum light, the influence of the progenitor structure, and in particular whether the progenitor is compact or extended, has abated so the spectral properties are primarily sensitive to composition (\nifs\ and other elements). The mixing of \nifs\ is also less important since the $\gamma$-ray mean free path is longer, even for weak mixing \citep{D15_SNIbc_I}. The full ejecta is turning optically-thin in the continuum so that the spectrum forms over the full ejecta, allowing a probe of most of the ejecta mass (the outermost fast expanding layers of the ejecta are optically thin at this time but they contain very little mass and thus bear less significance for understanding the progenitor star composition and mass).

\subsection{Differences with previous simulations of SNe I\lowercase{bc} with \cmfgen}

Simulations of SNe Ibc have been performed in the past with \cmfgen\ \citep{dessart_11_wr, d12_snibc,D15_SNIbc_I,D16_SNIbc_II,dessart_98bw_17} and so it might be useful to clarify what distinguishes these various studies.

In the first four studies, we used a selection of binary progenitor models from \citet{yoon_ibc_10}. In \citet{dessart_11_wr}, the selection was augmented to include single-star models. In \citet{dessart_98bw_17}, only single star models were included (all based on a 40\,\msun\ progenitor star, often with allowance for rotation), and the study was more focused on highly energetic explosions leading to SNe Ic-BL and GRB/SNe Ic. In the present work, we start with He star models, which most likely arise from binary evolution. In contrast, the models from \citet{yoon_ibc_10} were evolved consistently from the ZAMS in a binary system. Our progenitor models were computed with \mesa\ while those of  \citet{yoon_ibc_10}  were computed with the code \bec\ and \kepler. Finally, the mass loss prescriptions differ significantly for the WR stage. In \citet{yoon_ibc_10}, the mass loss rate calibrations were taken from \citet{hamann_mdot_82} and \citet{hamann_wr_95}, but scaled down by a factor 5 or 10, which tended to produce massive WC stars at core collapse. In the present study, the WR mass loss rates are greater (see Section~\ref{sect_prog} and \citealt{yoon_wr_17}). Furthermore, in the models of \citet{yoon_ibc_10}, WR mass loss rate were a function of luminosity (and hydrogen mass fraction, which is irrelevant for the present He-star models) and did not consider the difference between WN and WC/WO stars. Therefore the amount of He at the pre-SN stage was a smooth function of the progenitor mass, in contrast with the progenitor models produced in this work.

The treatment of the explosion in these studies is comparable, irrespective whether the code used was \kepler\ or \v1d. A small or moderate-size nuclear network was employed (accounting for 21 up to 71 isotopes). The explosions were generated artificially, either with a piston or a thermal bomb. Given the focus of our present and past studies, only the total explosion energy and \nifs\ mass matter.

\begin{figure*}
\includegraphics[width=0.5\hsize]{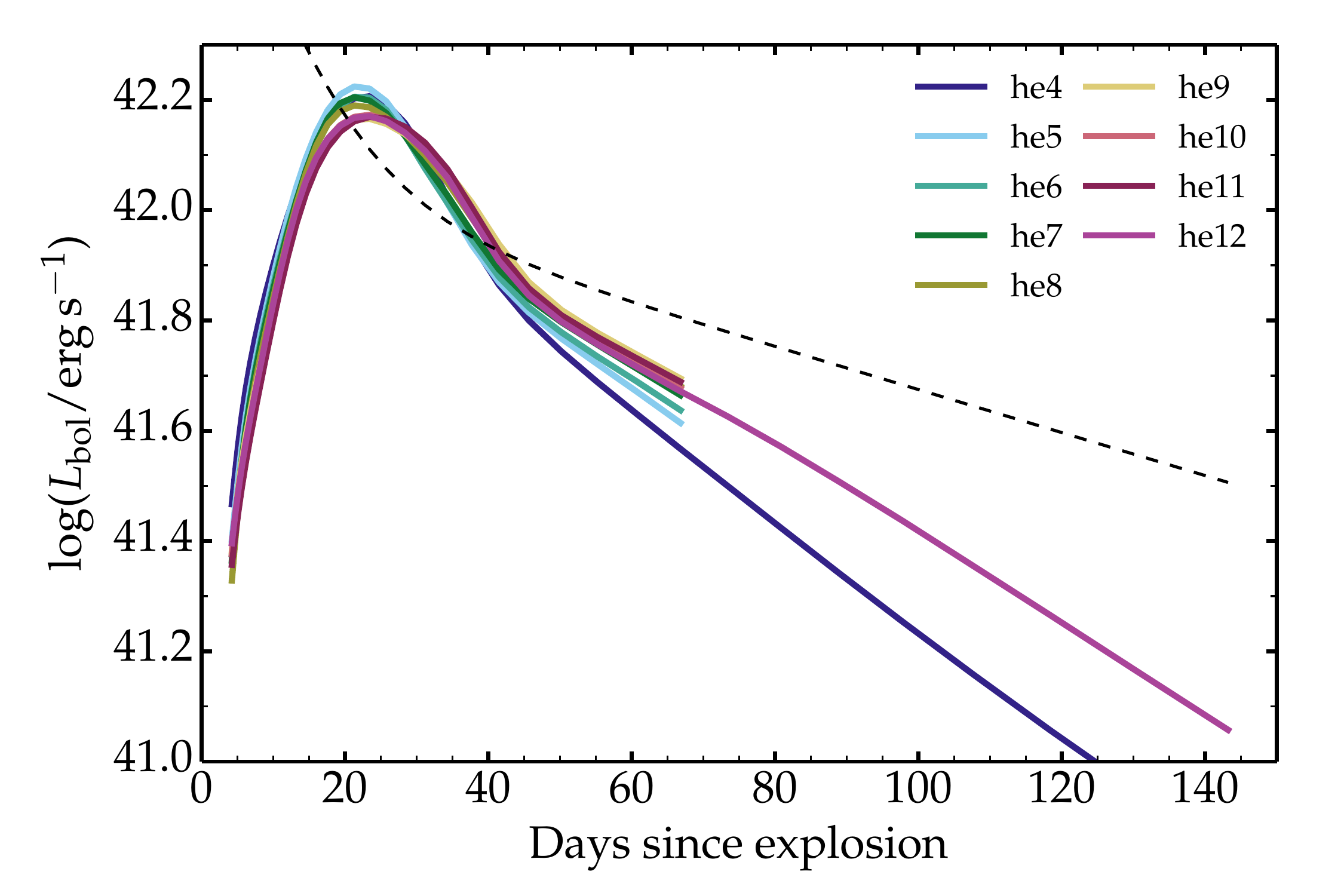}
\includegraphics[width=0.5\hsize]{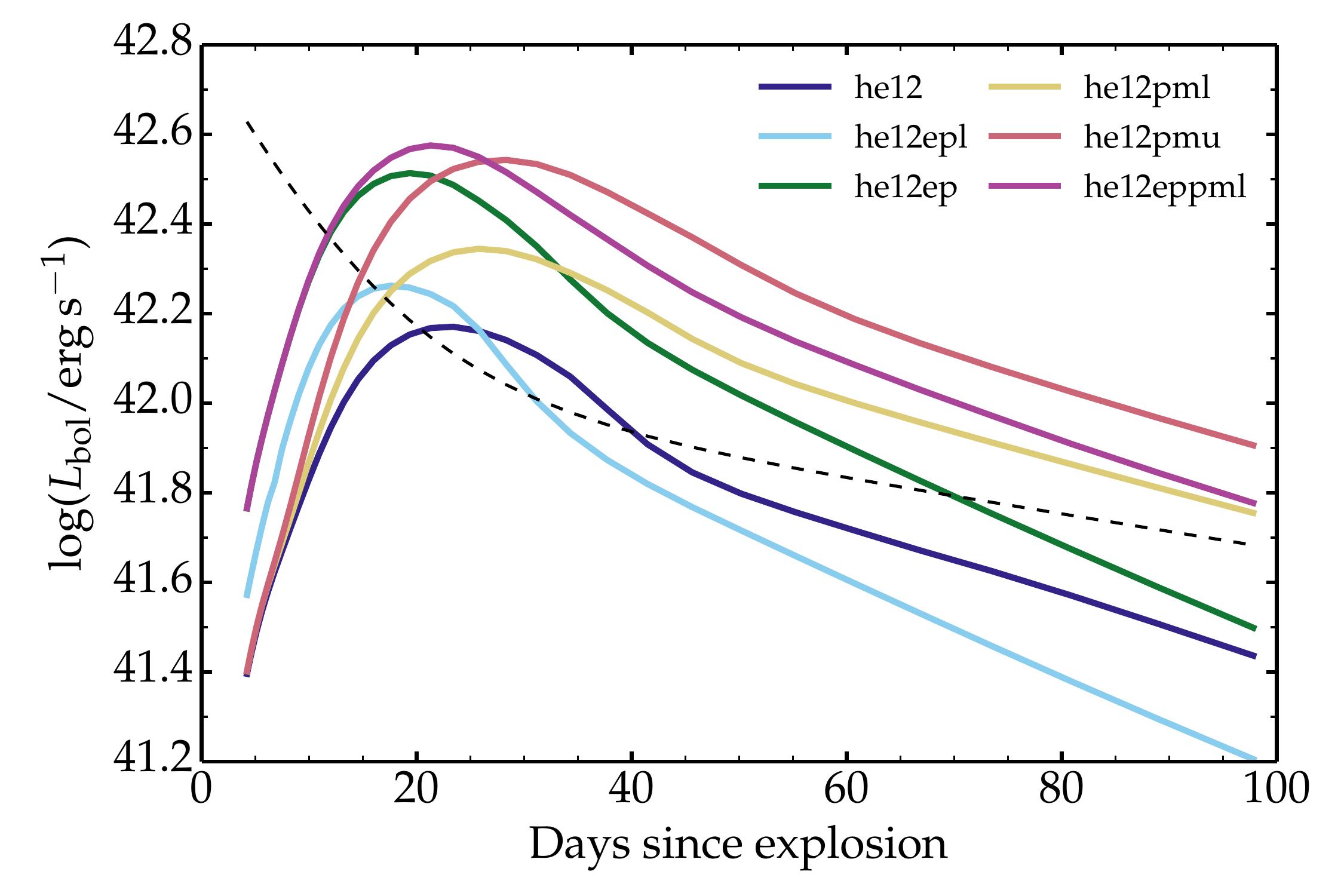}
\caption{{\it Left:} Bolometric light curves computed with \cmfgen\ for the explosion models based on He-star models he4 to he12. All nine models have the same \nifs\ mass of 0.08\,\msun\ and the same ejecta kinetic energy per unit mass of $5 \times 10^{50}$\,erg\,$M_\odot^{-1}$. {\it Right:} bolometric light curves for model he12 and variants that include power from the compact remnant (he12pml and he12pmu), a higher kinetic energy (he12epl), a higher kinetic energy and higher \nifs\ mass (he12ep), and both combined (he12eppml). Details on these ejecta model parameters are given in Table~\ref{tab_ejecta}. In both panels, the dashed line gives the instantaneous decay power emitted by an initial \nifs\ mass of 0.08\,\msun.
\label{fig_lbol}
}
\end{figure*}

An important difference between these studies is the treatment of non-thermal processes in \cmfgen. These were ignored in \citet{dessart_11_wr}, so that model spectra were only accurate as long as \nifs\ decay did not influence the spectrum formation region. This prevented the discussion of the model spectra after about a week past explosion. From 2012 onwards, non-thermal processes were treated so that the non-LTE physics was accurately handled (for example with respect to the presence or absence of He\one\ lines). The treatment of radioactive decay was flexible from the start. We have been using local (when the $\gamma$-ray mean free path is very small) or non-local energy deposition. For the latter, a Monte-Carlo technique was used until 2016 to transport $\gamma$-rays. Since then, we have been solving the grey pure-absorption radiative-transfer equation for the $\gamma$-rays. The Monte Carlo method is a little more accurate but the second method is more convenient and faster (see discussion in \citealt{wilk_gammaray_19}).

Based on our experience, the decisive difference between this work and previous studies of SNe Ibc with \cmfgen\ is the use of a new set of progenitor models. Specifically, it is the revision of WR mass loss rates affecting our He star models, most likely produced from binary evolution, that brings the possibility of explaining the properties of standard SNe Ibc. This is the central motivation for the present work.

\section{Results: light curve properties}
\label{sect_lc}

Figure~\ref{fig_lbol} shows the bolometric light curve for a subset of our models. Not all models are shown since there is a lot of degeneracy, but the main light curve characteristics are given in Table~\ref{tab_sn_rad}. In this table, we provide the rise time to maximum, the $UVOIR$ luminosity or $V$-band magnitude at maximum, and the brightness decline between the time of maximum and 15 days later.

The left panel of Fig.~\ref{fig_lbol} shows that the choice of a fixed value of $5 \times 10^{50}$\,erg\,$M_\odot^{-1}$ for the ejecta models he4\,$-$\,he12 yields essentially the same light curve properties up to about 40\,d after explosion. This emphasizes again the degeneracy of SN Ibc light curves and the difficulty of constraining the ejecta mass or kinetic energy. The rise time to bolometric maximum ranges between 18.2 and 24.4\,d. The rise time to $V$-band maximum occurs a little earlier and is between 17.8 and 22.4\,d. The bolometric luminosity peaks between 1.46 and $1.68 \times 10^{42}$\,\ergs\ (corresponding to the range 42.16 to 42.2 in the log). Lighter models tend to peak earlier, at a larger luminosity, and decline faster both early after maximum as well as at all times in the nebular phase. These differences are however small.

\citet{drout_11_ibc} found that their sample of SNe Ib and Ic had light curve properties that were statistically undistinguishable from each other. The left panel of Fig.~\ref{fig_lbol} shows that realistic stellar evolution and stellar explosion models may reproduce this property. Our two main assumptions, which are physically motivated and not particularly stringent, are that 1) WC stars at the origin of SNe Ic must have a larger mass loss rate than previously proposed \citep{yoon_wr_17}, and that 2) the explosion energy must increase with progenitor mass so that the $E_{\rm kin} / M_{\rm ej}$ must be comparable between Ib and Ic ejecta. With the mass loss rate prescription of \citet{yoon_wr_17}, the difference in final mass between model he4\,$-$\,he5 and he10\,$-$\,he12 is only of a factor two, meaning that one needs to invoke a factor of about two difference in explosion energy (ignoring differences in progenitor binding energy) between model he4\,$-$\,he5 and he10\,$-$\,he12 to yield similar light curve properties for each set. This factor would be reduced at higher metallicity since the final mass of models he10\,$-$\,he12 would be lower.

\input{table_prop_sn_radiation.tex}

In Type II SNe, a factor of ten is inferred for the explosion energy between the low luminosity and the high luminosity events so invoking a factor of two here does not seem unreasonable. Previously, with the smaller WC mass loss rates, the WC star models reached core collapse with a final mass around 10\,\msun\ (see extended discussion in \citealt{yoon_ibc_15}), which then requires a much greater range in explosion energy. Previous simulations for such high mass WR stars show that the rise times and light curve widths are large and significantly in conflict with observations \citep{ensman_woosley_88,dessart_11_wr,dessart_98bw_17}, unless one invokes an explosion energy of many 10$^{51}$\,erg (see, for example, \citealt{bersten_08D_13} for SN\,2008D). While there is still much uncertainty surrounding the core-collapse explosion mechanism, explosion energies in excess of $2 \times 10^{51}$\,erg are hard to justify routinely \citep{ertl_ibc_20} -- they must represent an exception rather than the norm. A sensible way out of this energy crisis is to invoke lower mass progenitors producing lower mass ejecta, since ejecta with a similar $E_{\rm kin} / M_{\rm ej}$ have a similar light curve.

The right panel of Fig.~\ref{fig_lbol} shows the bolometric light curves for model he12 and variants in which the explosion energy is increased, the \nifs\ mass is increased, or a magnetar power is introduced (various combinations of these are explored -- see Table~\ref{tab_ejecta}). The goal of this exploration is not to obtain a perfect fit to the observations but to investigate the range of light curves that can be produced through moderate changes in the properties of model he12. When magnetar power is invoked (or alternatively fallback accretion), it is possible with a modest energy (here we use one thousands of the magnetar power or energy that is required to explain SLSNe Ic; \citealt{KB10}) to double or triple the SN Ibc  luminosity during the photospheric phase.

\begin{figure*}
\includegraphics[width=\hsize]{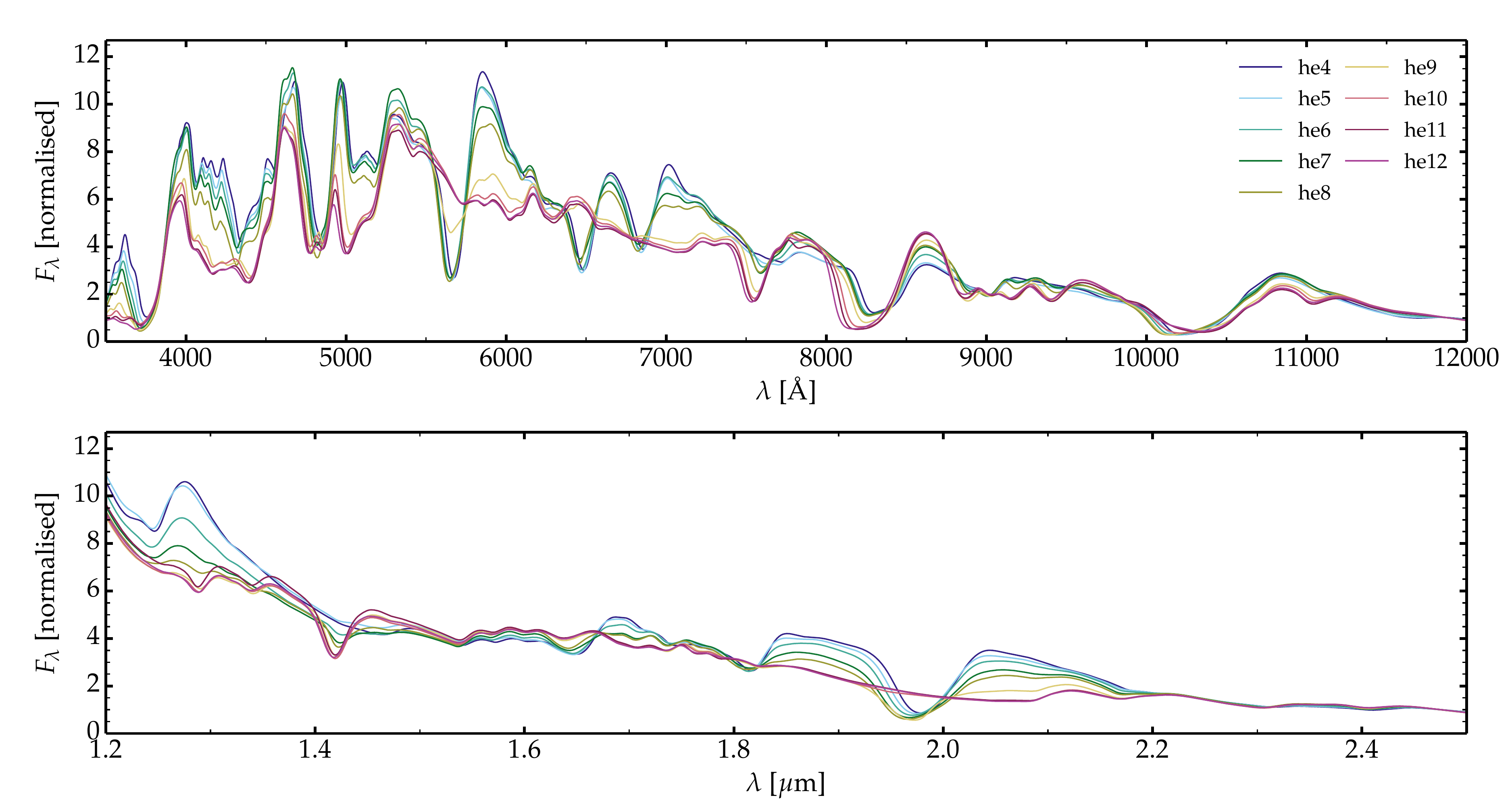}
\caption{Comparison of the optical (top) and near-infrared (bottom) spectra at bolometric maximum for the He-star explosion models he4\,$-$\,he12. There is a clear separation between models that show He\one\ lines (at 5876, 6678, 7065, 10830, 20581) and those that do not. Model he9 is an intermediate case. The presence or absence of He\one\ lines is easy to see in the optical. He\one\,10830\,\AA\ overlaps with O and Mg lines so a feature is seen in this spectral region irrespective of the He content. He\one\,\,20581\,\AA\ gives a clear signature. Note also that all models have about the same color at maximum (modulo the strength of lines). For He\one\ line identifications, see Fig.~\ref{fig_comp_flux_max_he}.
\label{fig_comp_flux_max}
}
\end{figure*}

An extended set of SNe IIb/Ib/Ic simulations computed with \cmfgen\ with similar physics was presented in \citet{D15_SNIbc_I,D16_SNIbc_II} so the various correlations and dependencies will not be repeated here. But we can say that an increase by a factor of two in \nifs\ mass or explosion energy, or the introduction of power from a $\sim$\,30\,ms period magnetar may explain the existence of SNe Ibc with a peak luminosity up to 10$^{42.6}$\,\ergs.  With the exception of a few outliers, this set of models encompasses the whole range of peak luminosities of observed SNe Ibc \citep{lyman_ibc_16,prentice_ibc_16}. Outliers (i.e., SNe Ibc with a peak luminosity above 10$^{42.6}$\,\ergs) are few and will require more special circumstances. Because they require an extraordinary power source, these outliers may not need to be considered as part of the sample of more standard SNe Ibc.

\begin{figure*}
\includegraphics[width=0.5\hsize]{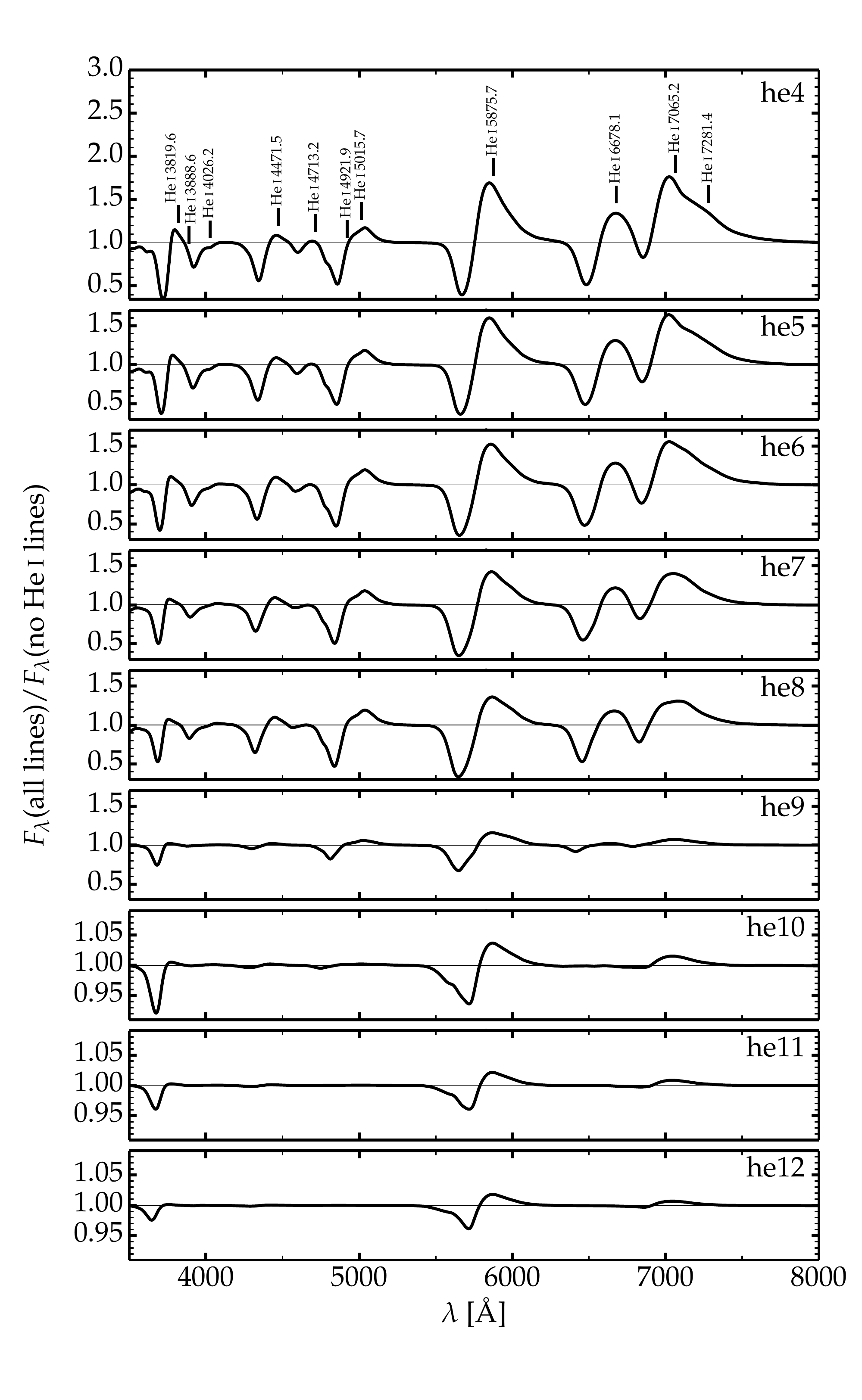}
\includegraphics[width=0.5\hsize]{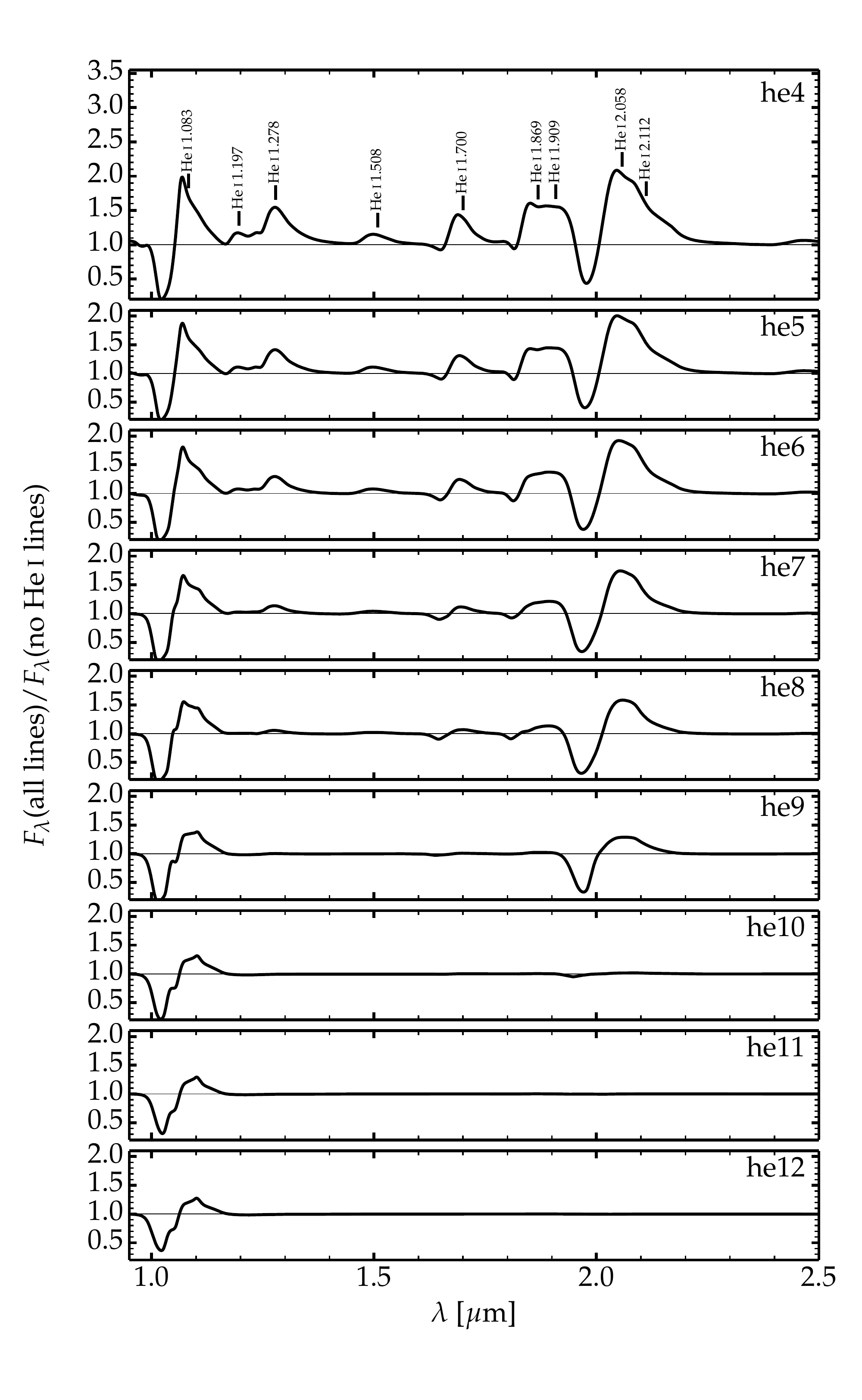}
\caption{Illustration of the He\one\ lines in our model sequence he4\,$-$\,he12 at bolometric maximum. The quantity shown is the ratio of the total flux (i.e. computed by accounting for all bound-bound transitions of all species) with the flux obtained by excluding He\one\ bound-bound transitions. In models he4 to he8,  He\one\ lines are not limited to 5876, 6678, and 7065\,\AA\ but are instead present throughout the optical up to 7281.4\,\AA. Note also that their strength diminishes as we progress from model he4 to he8, and progressively vanish as we progress from model he9 to he12. The ordinate scale is not kept the same for all panels, to better reveal the strong variation in the strength of He\one\ lines between models. While He\one\ optical lines stand 30 to 60\% above the continuum in models he4\,$-$\,he8, their strength is at the 1\% level in models he10\,$-$\,he12 (such weak lines are hard to detect without high $S/N$).
\label{fig_comp_flux_max_he}
}
\end{figure*}

\section{Results: spectral properties}
\label{sect_spec}

Progenitor models he4 to he12 exhibit a monotonic drop in the mass of the He-rich shell and surface He mass fraction (Fig.~\ref{fig_prog}). Model he9 is at the transition between models he4\,$-$\,he8 that have retained a He-rich shell (the shell where He has a 98\,\% mass fraction) and models he10\,$-$\,he12 that have only a low-mass outer shell with some residual He on top of a growing CO core (see Fig.~\ref{fig_prog} and Table~\ref{tab_prog}). These properties led \citet{yoon_wr_17} to propose that ejecta models he4\,$-$\,he8 would produce SNe Ib and the rest SNe Ic.

Figure~\ref{fig_comp_flux_max} shows the optical and near-infrared spectra for model he4 to he12 at the time of bolometric maximum. This epoch is unambiguously defined and corresponds roughly to the time of spectral classification for many SNe Ibc observed to date. We see that models he4 to he8 have very similar spectral energy distributions, with the obvious presence of He\one\,5876\,\AA. This line is half as strong in model he9, and very weak in models he10\,$-$\,he12. Hence, we see that there is a clear dichotomy for the presence of He\one\,5876\,\AA, being present in all models with a massive He-rich shell (wherein the He mass fraction is essentially equal to $1-Z$; models he4 to he8) and being absent in models with a truncated or absent He-rich shell (wherein He is part of a mixture with C, O, Ne, and Mg) .

The He\one\ line strengths are controlled by complex, non-linear, non-LTE effects, and strongly influenced by non-thermal processes (see extensive discussion in \citealt{d12_snibc} and \citealt{li_etal_12_nonte}). Despite this complexity, one can surmise that the progressive drop in the strength of He\one\ lines is related to the progressive decline in He abundance (concomitant with the progressive increase in O abundance) in the spectrum formation region (recall that the \nifs\ abundance profile in velocity space is similar in models he4\,$-$\,he12 so the mixing is effectively similar in this model set). This region extends from below the photosphere up to large velocities, and is thus not confined to the photosphere. However, the photosphere composition gives a clue as to why the He\one\ lines become so weak in models he10 to he12.  In table~\ref{tab_sn_rad}, we give the He and O mass fraction at the Rosseland-mean photosphere at the time of maximum. While the He mass fraction stays above $\sim$\,0.6 in model he4\,$-$\,he8 (it dominates the composition), it drops below 0.2 in models he10\,$-$\,he12 (it is below 0.2 nearly throughout the ejecta in those models; see Fig.~\ref{fig_prog}). As already discussed in the introduction, the lower He abundance implies that He captures a lower share of the total decay power absorbed by the plasma. This is aggravated by the fact that the ejecta masses increase as we progress from model he4 to he12, while the powering source is kept the same (the \nifs\ mass is the same in this model set and equal to 0.08\,\msun). Since the \nifs\ abundance profile is the same in models he4\,$-$\,he12, the chemical stratification of our He star models implies that the \nifs\ abundance is greater where He is present in lower-mass He-star models (i.e., model he4 relative to he12). Finally, a lower He abundance implies smaller He level populations, which would lower the strength of He\one\ lines.

Figure~\ref{fig_comp_flux_max} shows that in the near infrared, the He\one\,1.083\,$\mu$m is a poor discriminant for He abundance since it shows a similar strength in all models he4 to he12 (this stems from the overlap with multiplet lines due to C\one, O\one, and Mg\two; \citealt{D15_SNIbc_I}), although the absorption extends to a larger velocity in He-rich models, as expected. In contrast, we find that the He\one\,2.0581\,$\mu$m line is an excellent tracer for the presence of a massive He-rich shell in the progenitor, just like He\one\,5876\,\AA\ in the optical range.

In general, the optical He\one\ lines that are studied in SNe Ib are limited to those at 5876, 6678, and 7065\,\AA. In practice, numerous other He\one\ lines are present. Figure~\ref{fig_comp_flux_max_he} shows model spectra for the sequence he4 to he12 at the time of bolometric maximum (same time as in Fig.~\ref{fig_comp_flux_max}), and more specifically the ratio of the total flux to that obtained by excluding bound-bound transitions associated with He\one. This ratio therefore reflects the contribution (akin to an equivalent width wherein the ratio is taken between the total flux and the continuum flux) due to He\one\ lines. The strength of He\one\ lines  decreases steadily as we progress through the sequence from ejecta model he4 to ejecta model he12. The maximum He\one\ line flux stands $\sim$\,50\% above the continuum flux in model he4, but drops to only $\sim$\,1\% above the continuum in model he12 (level at which it is hard to detect).

In the optical, the strongest He\one\ lines are at 5876 and 7065\,\AA\ (the latter is very broad because it overlaps with He\one\,7281.4\,\AA), followed by He\one\,6678\,\AA. However, there are numerous He\one\ lines in the blue part of the optical, and specifically at  5016 and 4922\,\AA\ (these overlap with Fe\two\ lines and may thus be un-noticed), 4713, 4471, 4026, 3889, and 3820\,\AA\ (the last two overlap with Ca\two\,H\&K). Other He\one\ lines in the optical are weaker and thus not discussed here. The He\one\ lines at 5016 and 4922\,\AA\ are at the origin of the difference observed in the 4900\,\AA\ region between the mean spectra of SNe Ib and SNe Ic presented by \citet{liu_snibc_15}. In the near infrared, besides He\one\ lines at 1.083 and 2.058\,$\mu$m, there are relatively strong He\one\ lines at 1.278, 1.70, 1.869, 1.909, and 2.112\,$\mu$m (the latter overlaps with He\one\,2.0581$\mu$m). As visible in Fig.~\ref{fig_comp_flux_max}, there is a clear similarity of spectral properties with respect to He\one\ lines between models he4 to he8 on the one hand, and model he10 to he12 on the other hand. Model he9 is an intermediate case where the SN classification may be ambiguous.

We see from these simulations that models with a He-rich shell exhibit optical spectra that are nearly exclusively composed of He\one\ lines (see also early-time models in \citealt{dessart_11_wr}). Classifying a SN as Type Ib is trivial in this case. But it also means that whenever there is ambiguity about the classification as Ib or Ic, the progenitor star cannot have a He-rich shell and is most likely poor in He. Here, in models he10\,$-$\,he12, the total mass of He is less than about 0.3\,\msun. With our He-star models, we see a direct correspondence between the WR classification as type WN or WC, and the SN classification as Type Ib or Ic.

\begin{table*}
\caption{
Characteristics of our selected sample of Type Ib and Ic SNe.
\label{tab_obs}
}
\begin{center}
\begin{tabular}{
l@{\hspace{4mm}}c@{\hspace{4mm}}c@{\hspace{4mm}}
c@{\hspace{4mm}}c@{\hspace{4mm}}c@{\hspace{4mm}}
c@{\hspace{4mm}}c@{\hspace{4mm}}c@{\hspace{4mm}}
c@{\hspace{4mm}}c@{\hspace{4mm}}c@{\hspace{4mm}}
c@{\hspace{4mm}}c@{\hspace{4mm}}
}
\hline
        SN       &  Type &    $D$      &  $\mu$   & $t_{\rm expl}$   &    $E(B-V)_{\rm gal}$    &  $R(V)_{\rm gal}$    &  $E(B-V)_{\rm host}$    &  $R(V)_{\rm host}$    &$z$    & Ref.\\
                     &  &     [Mpc]  &     [mag]  &     MJD [d]           &     [mag]        &     &   [mag]    &           &        &    \\
\hline
  sn2004gq &   Ib & 25.1 &  32.00 & 53346.9 &   0.06 &    3.1 &   0.08 &    3.1 &   0.0065  & CfA, CSP  \\
  sn2004gv &   Ib & 79.6 &  34.50 & 53345.3 &   0.03 &    3.1 &   0.03 &    3.1 &   0.0200  &  CSP \\
  sn2005hg &   Ib & 86.0 &  34.67 & 53665.8 &   0.09 &    3.1 &   0.00 &    3.1 &   0.0210  & CSP   \\
  iPTF13bvn &  Ib &  25.5 &  32.04 & 56459.2 &   0.04 &    3.1 &   0.17 &    3.1 &   0.0045 & $a$ \\
   sn2008D &   Ib & 31.0 &  32.46 & 54474.0 &   0.02 &    3.1 &   0.60 &    3.1 &   0.0065 & $b$ \\
\hline
   sn2009jf &  Ibc$^g$ &  33.9 &  32.65 & 55099.5 &   0.05 &    3.1 &   0.11 &    3.1 &   0.0079   & $c$ \\
  sn2017ein &  Ibc$^g$ &  20.0 &  31.51 & 57898.5 &   0.02 &    3.1 &   0.38 &    3.1 &   0.0027  & $d$ \\
\hline
  sn2007gr &   Ic &  9.3 &  29.84 & 54325.5 &   0.06 &    3.1 &   0.03 &    3.1 &   0.0017 & $e$ \\
 sn2004aw &  Ic &  68.2 &  34.17 & 53073.0 &   0.02 &    3.1 &   0.35 &    3.1 &   0.0160 & $f$  \\
\hline
\end{tabular}
\end{center}
{\bf Notes:} In the following, CfA refers to \citet{bianco_etal_14} and CSP to \citet{taddia_ibc_lc_18}. $a$: \citet{bersten_iPTF13bvn_14}; $b$: \citet{modjaz_08D}; $c$: \citet{valenti_09jf_11}; $d$: \citet[we take the average of their values for the distance and the extinction]{vandyck_2017ein_18}; $e$: \citet{hunter_07gr}; $f$: \citet{taubenberger_04aw_06}. $g$: the classification as Ibc arises because of the use of the intermediate model he9 at the junction between models that are unambiguously associated with SNe Ib (he4\,$-$\,he8) and SNe Ic (he10\,$-$\,he12).
\end{table*}

\section{Comparison to observations}
\label{sect_comp_obs}

\subsection{Observational data}
\label{sect_obs}

We selected a small but representative sample of well observed SNe Ib and Ic. One criterion was that both optical and near-infrared photometry was available from pre-maximum to the nebular phase (in cases where the near-infrared photometry is lacking, we only compare the post-maximum spectrum). Secondly, a good quality spectrum was required around the time soon after maximum.  At this epoch, the variations in progenitor radius no longer influence the SN radiation, departures from spherical symmetry might be weaker, and the spectrum formation region is extended so that most of the ejecta (in terms of mass rather than velocity since the outermost layers traveling at high speed are optically thin) is probed. Hence, at this epoch, the SN spectrum is primarily sensitive to the composition (He, CNO, intermediate mass elements, iron group elements) and the heating source (which provides the source of radiation, controls the ionization etc). Finally,  all data had to be public.

\begin{figure*}
\begin{center}
\includegraphics[width=0.49\hsize]{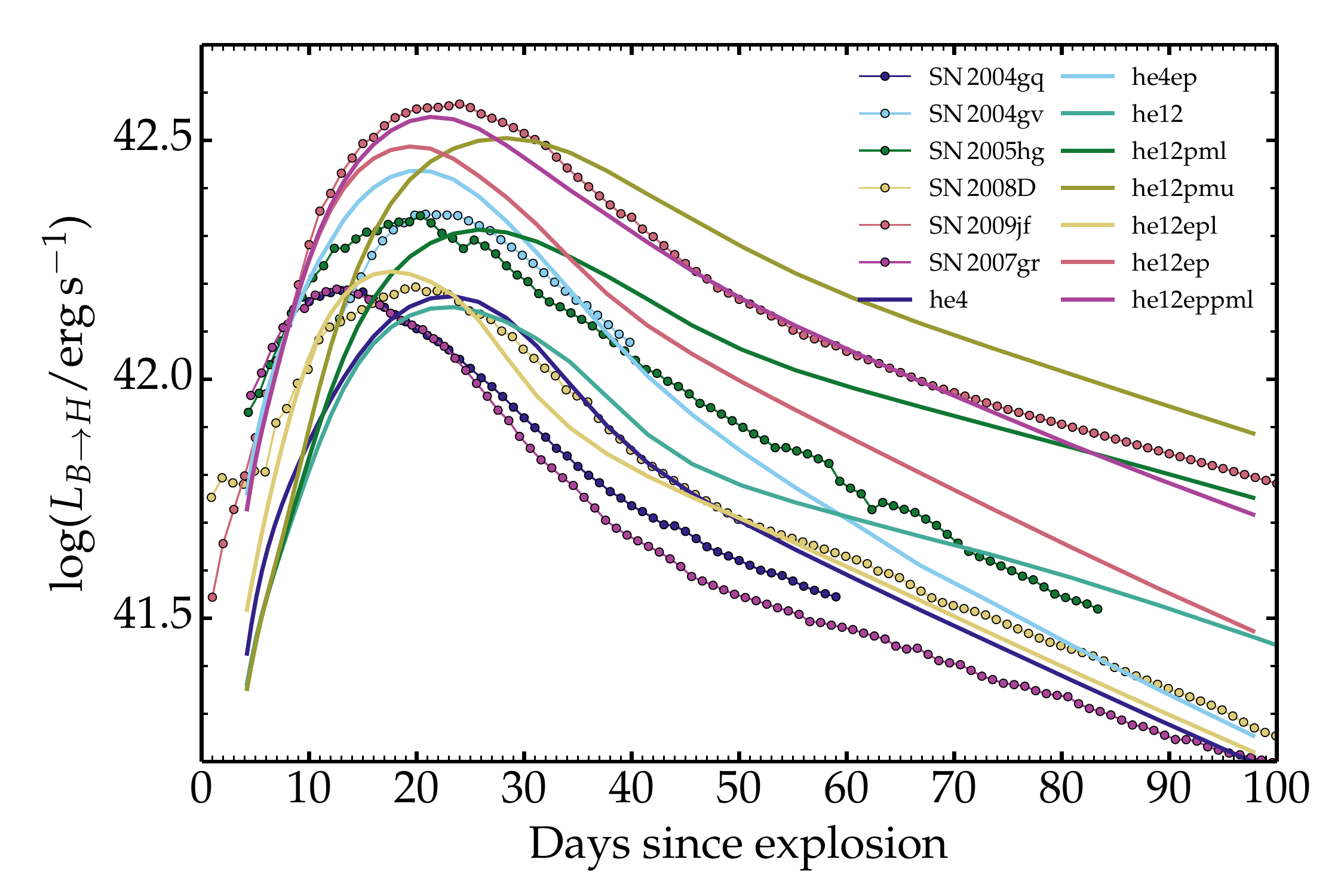}
\includegraphics[width=0.49\hsize]{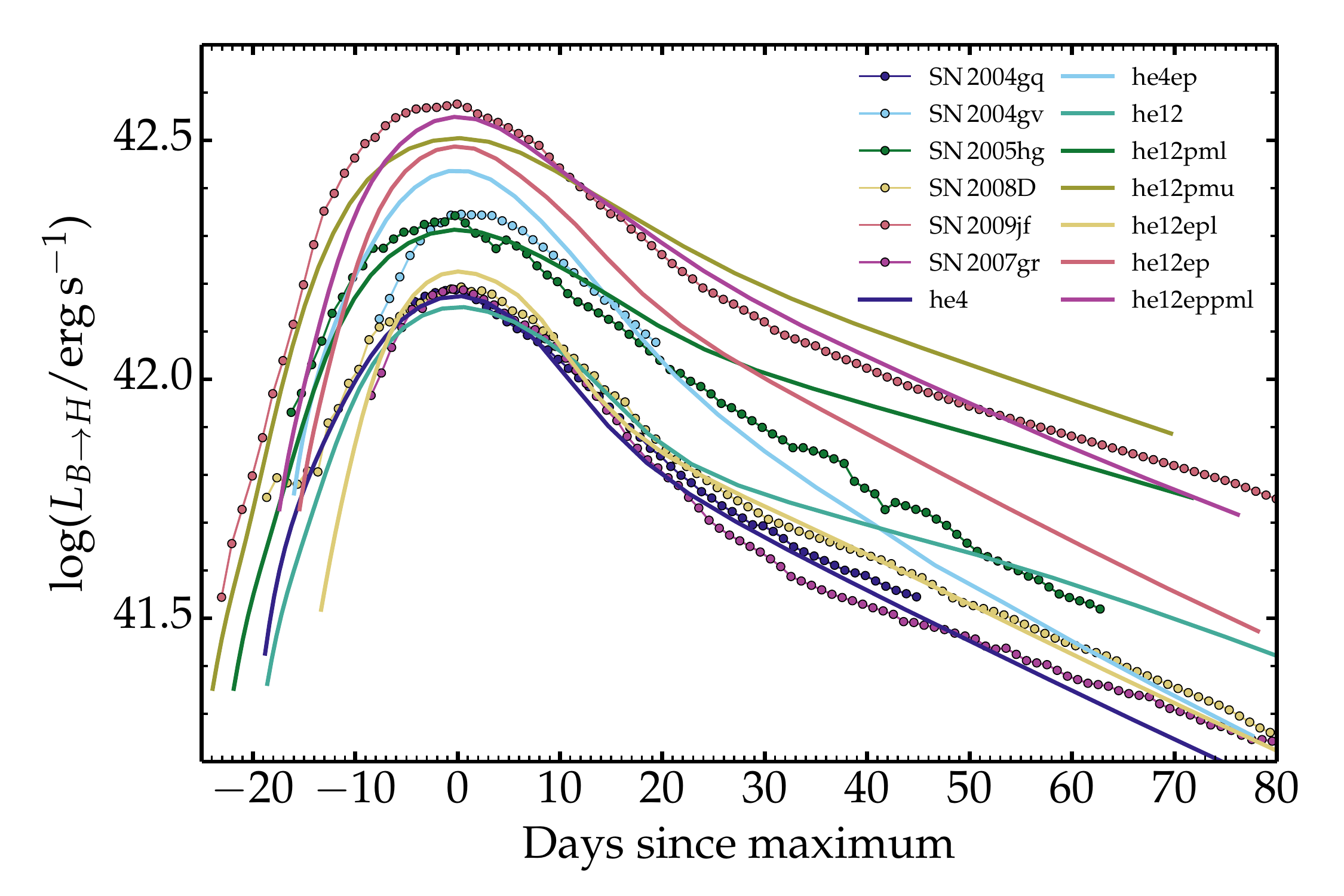}
\caption{Evolution of the luminosity falling within the $B$ and $H$ bandpasses for a sample of SNe Ibc and a representative sample of models taken from this study. The time origin is the time of explosion (left) or the time of maximum luminosity (right).
\label{fig_comp_lum}
}
\end{center}
\end{figure*}

The sample of observed SNe that we selected is presented in Table~\ref{tab_obs}, together with their characteristics (distance, reddening, redshift) and corresponding references. The sample of SNe Ib includes SNe\,2004gq, 2004gv, 2005hg, iPTF13bvn, and SN\,2008D. The sample of SNe Ic (or cases that could be considered intermediate between Ib and Ic, although these are markedly different from the selected SN Ib sample) includes SNe\,2009jf, 2017ein, 2007gr, and 2004aw.

Using the observational characteristics of each selected SN, we convert from photometry to flux and build the SN luminosity falling between the $B$-band and the $H$-band. In most cases, the values we infer for the peak luminosity is similar to previous similar inferences \citep{lyman_ibc_16,prentice_ibc_16,meza_anderson_ni56_20}, although in many cases (including SNe that we do not include in our sample),  reddening is uncertain (in some cases, literature values seem to be in error). This is particularly relevant because, perhaps because of their association with denser and younger stellar populations, many SNe Ibc are affected by a very large inferred reddening, a feature that does not seem so prevalent with SNe II-P.  Intriguingly, many SNe Ibc with large inferred peak luminosities (around or above 10$^{42.6}$\,\ergs) also have a large inferred reddening, so one wonders whether these hard-to-explain luminosities result from an overestimate in the reddening. Finally, the same photometry-to-flux conversion procedure is applied to the photometry of all our models so we can directly compare quasi-bolometric light curves for our sample of observed SNe Ibc and our models (there is thus no concern about the missing flux below and beyond the wavelength range considered). We have tested our procedure on models: taking the flux from \cmfgen, we compute the photometry from $U$ to $K$, and then proceed back to infer the luminosity falling between $U$ and $K$. This gives the same result at the percent level as that obtained by computing this luminosity directly from the original \cmfgen\ spectrum.

\begin{figure*}
\begin{center}
\includegraphics[width=0.8\hsize]{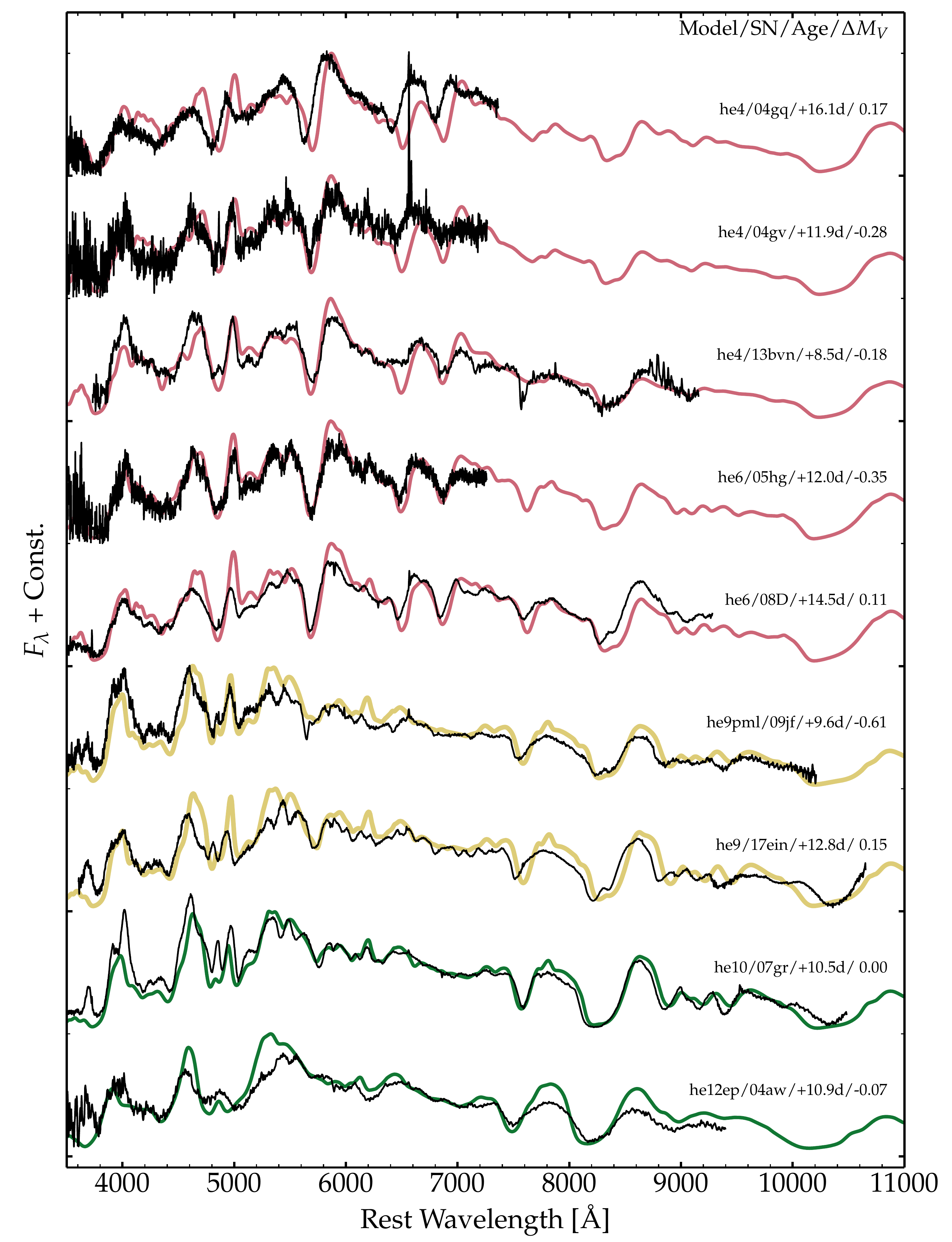}
\caption{Comparison of observed optical spectra (corrected for redshift and reddening; see Table~\ref{tab_obs}) for a sample of SNe Ib, and Ic with a selection of models from this study. Colors red, yellow, and green correspond to type Ib, Ibc, and Ic SN models.
\label{fig_comp_spec}
}
\end{center}
\end{figure*}

\subsection{Light curve comparisons}
\label{sect_obs_lc}

Figure~\ref{fig_comp_lum} compares the $B$ to $H$ luminosity evolution for our sample of SNe Ib and Ic (we only include the observed SNe that have photometry covering $B$ to $H$) with a limited set of our models. As evident from Fig.~\ref{fig_lbol}, the ejecta mass has a weak influence on our model light curves during the photospheric phase because by design we adopted a fixed value of $E_{\rm kin} / M_{\rm ej}$ for the model sequence he4 to he12. The selection of models shown in Fig.~\ref{fig_comp_lum} is thus limited to he4, he4ep, he12, he12epl, he12ep (he4 and he12 on the hand, and he4ep and he12ep appear similar; he12epl is the counterpart to he12ep but with the same \nifs\ mass as he12), and then some variants of he12 with magnetar power (he12pml, he12pmu, he12eppml) -- see Table~\ref{tab_ejecta} for details.

In the left panel of Fig.~\ref{fig_comp_lum}, the light curves are shown with respect to the time of explosion (or rather the inferred time of explosion for the observations). One sees that the models encompass the range of peak luminosities between (relatively) fainter events like SN\,2008D and brighter events like SN\,2009jf. None of the models fit perfectly but it is clear that by invoking various combinations of the present range of ejecta masses, kinetic energies, \nifs\ mass, and a moderate power from the compact remnant, one can cover the parameter space occupied by the majority of SNe Ibc observations. The models show a wide range of decline rates at nebular times : modulations in $\gamma$-ray escape can lead to a fast decline, while at the opposite extreme, introduction of magnetar (or fallback-accretion) power leads to a slower fading at late times (as would also occur in the case of full trapping of $\gamma$-rays from \cofs\ decay).

For a number of models, there is a discrepancy in the rise time. Most of our models peak around 21\,d after explosion (see Table~\ref{tab_ejecta}; one exception is model he12epl, characterized by a large $E_{\rm kin}$ of $4.8 \times 10^{51}$\,erg but a modest \nifs\ mass of 0.08\,\msun) while numerous observed SNe Ibc have an inferred time of explosion between about 10 and 15\,d. If real, this short rise time requires models with a larger $E_{\rm kin} / M_{\rm ej}$ (as in model he12epl). Since the explosion energy applied in our models (for example he4ep, he12epl, or he12ep) is already large, in tension with what the explosion mechanism can allow (see \citealt{ertl_ibc_20}), one way to explain such short rise times is to invoke smaller ejecta masses, probably in the range 0.5-1\,\msun. In our sample, this applies especially to the Type Ib SN\,2004gq (which could arise from a low mass He giant; see, for example, \citealt{eldridge_13bvn_15}, \citealt{d18_ext_ccsn}) and to the Type Ic SN\,SN2007gr (which could arise from Case BB mass transfer in a tight binary; \citealt{dewi_bin_02,dewi_pols_03,tauris_ulstr_15}). The latter scenario was proposed to explain the observations of the fast rising and fast declining Type Ic SN\,1994I \citep{nomoto_94I_94}.

When plotting the same $B$ to $H$ luminosity evolution but now with respect to the time of maximum (right panel of Fig.~\ref{fig_comp_lum}), the agreement between observations and models is much improved. The light curves of SNe\,2004gq (Type Ib) and 2007gr (Type Ic), which nearly overlap, are well matched by model he4 and he12, confirming the proximity of SNe Ib and Ic light curves despite their distinct SN type \citep{drout_11_ibc}. This suggests that the main source of discrepancy here is the ejecta mass (for the fast rising SNe) while SNe Ibc with slightly higher peak luminosities may be explained by a weak power contribution from the compact remnant.

The existence of even more luminous SNe Ibc presents the same theoretical challenge that pertains to the existence of GRB/SNe Ic, SNe Ic-BL, or super-luminous SNe Ic. Luminous SNe Ibc may be considered transition objects between standard SNe Ibc and SNe Ic-BL or GRB/SNe. The theory explaining the latter would apply at a more moderate level in those transition objects.

 In \citet{prentice_ibc_16},  SNe Ibc (excluding GRB/SNe, SNe Ic-BL, and SNe IIb) with a peak luminosity greater than 10$^{42.6}$\,\ergs\ are SN\,2007C, SN\,2007uy, SN\,2009er, SN\,2011bm, PTF\,11rka, and PTF\,12gzk. Let us inspect these peak luminosities in more detail. For SN\,2007C, \citet{prentice_ibc_16} use a total reddening $E(B-V)$ of 0.77\,mag and a distance of 26.9\,Mpc. However, \citet{meza_anderson_ni56_20} use a total reddening that is much lower, with $E(B-V)$ of 0.47\,mag. This implies a 0.37 dex lower peak luminosity, bringing SN\,2007C much below 10$^{42.6}$\,\ergs. For SN\,2007uy, the inferred reddening is large with  $E(B-V)$ of 0.65\,mag, yielding a peak luminosity of 10$^{42.67}$\,\ergs\ \citep{prentice_ibc_16} or  10$^{42.58}$\,\ergs\ \citep{meza_anderson_ni56_20}. However, comparing the maximum light spectrum of SN\,2007uy to that of SN\,2008D (which has a similar inferred reddening), one sees that SN\,2007uy is much bluer, meaning it is likely not as reddened. Hence, just like for SN\,2007C, the reddening of 0.65\,mag used for SN\,2007uy is likely overestimated and therefore the peak luminosity of that SN is probably smaller, bringing it below  10$^{42.6}$\,\ergs. SN\,2009er is classified as a Type Ib-pec by \citet{bianco_etal_14} so not a standard SN Ib nor Ic.  SN\,2011bm and PTF\,11rka possess very broad light curves and large sustained luminosities at nebular times. Perhaps these are intermediate events requiring a magnetar (or whatever extra power source), and standing between standard SNe Ibc and objects like SNe 2005bf \citep{tominaga_05bf_05,folatelli_05bf_06,maeda_05bf_07} and SLSNe Ic \citep{pasto_10gx_10,quimby_slsnic_11}. Finally, PTF\,12gzk shares similar properties to SNe Ic-BL and is thus not a standard SN Ic \citep{benami_12gzk_12}. The evolutionary paths and metallicities leading to GRB/SNe and SLSNe Ic are distinct from those relevant to the majority of SNe Ibc \citep{aguilera_evol_ic_18}. We thus see that an overestimate in the reddening is in part responsible for the large inferred peak luminosities (the error applies to the whole light curve) of several SNe Ibc that have so far been considered outliers in the community.

So, from this exploration, we see that all SNe Ibc having a peak luminosity above  10$^{42.6}$\,\ergs\ are either peculiar events, extraordinary events like SNe Ic-BL or GRB/SNe, or were given an overestimated reddening.  We can draw a parallel with SNe II. What would be the distribution of SNe II peak luminosities (or luminosities at 10\,d after explosion or discovery, since most SNe II do no have a clear peak in their bolometric light curve) if we included all H-rich events? We would find low luminosity events (the faint SNe II-P), ``standard"-luminosity events (SNe II-P and SNe-pec, powered by shock deposited energy prior to shock breakout, and \nifs\ decay power), and then very  luminous events (all SNe IIn, in which the power arises from interaction with CSM). Explaining this whole range with a \nifs-power model combined with the standard neutrino-driven mechanism would lead to an inconsistency. This inconsistency may arise because high luminosity events are not powered by \nifs. The same applies to the SN Ibc sample, for which one finds that the \nifs-power model combined with the standard neutrino-driven mechanism cannot explain the high-luminosity end. These brighter events probably owe their luminosity to a distinct power source.

It thus seems that the grid of models we
present here, based on moderate explosion energies, a range of \nifs\ masses between 0.08 and 0.16\,\msun, or boosted by a slowly rotating magnetar can reproduce the energetics of nearly all SNe Ibc light curves. There is no ``extreme" physics involved here. SNe Ibc that are more luminous at maximum and sometimes later require extra power from the compact remnant. It thus seems plausible that SNe Ibc arise from a standard core-collapse explosion mechanism, at best aided by a modest energy injection from the compact remnant \citep{ertl_ibc_20}. Such a modest extra power may also be present in Type II SNe but difficult to quantify because their ejecta are optically thick for much longer than in SNe Ibc.

\subsection{Spectral comparisons early after maximum light}
\label{sect_obs_spec}

An advantage of our modeling with \cmfgen\ is the computation of the time evolving spectral energy distribution, from which one can extract the bolometric light curve, photometric light curve in any bandpass, as well as the spectral evolution in any wavelength region between 100\,\AA\ and 100\,$\mu$m. In other words, the same numerical code yields the information on the photometric and spectroscopic evolution of a given model. Only few codes in the community use this approach, but with a different treatment of non-LTE processes and typically not applied to the modeling of SNe Ibc ({\tt SEDONA}, \citealt{sedona}; {\tt ARTIS}, \citealt{sim_07,kromer_sim_09}; {\tt JEKILL}, \citealt{jekill}). In contrast, SN studies generally use distinct codes with distinct physics to compute the  photometric and spectroscopic evolution, so that physical consistency is not guaranteed in those works. We have presented numerous studies of SNe Ibc in the past, and showed their full bolometric, multi-band photometric, color, and spectral evolution from early to late times \citep{dessart_11_wr,d12_snibc,D15_SNIbc_I,D16_SNIbc_II,dessart_98bw_17}.  Here, we focus on the spectral properties around the time of maximum.

Figure~\ref{fig_comp_spec} presents a comparison of optical spectra of SNe Ib and Ic with a selection of our models. The observed spectra are first corrected for redshift and reddening, and then normalized to the model spectra at a wavelength between 6000 and 7000\,\AA. This normalization is done to ease the comparison. In practice, the offset in absolute $V$-band magnitude between observations and models at the corresponding epoch is shown in the label at right (a negative value means that the observations are brighter than the model). The label also gives the SN name, the model name, and the time elapsed since $V$-band maximum for both. The color coding is used to group together the SNe Ib (models in red), the intermediate cases between SNe Ib and Ic (models in yellow), and the SNe Ic (models in green). Here, the grouping by SN type is not meant to reflect any specific equivalent-width estimate for the strength of putative He\one\ lines. Instead, we search for a global compatibility between observed and model spectral properties. Our goal is to check whether our grid of stellar evolution and explosion models mirrors the diversity of observed SNe Ib and Ic.

The SNe Ib that show unambiguously He\one\ lines are well fitted with models he4 and he6 (models he4 to he8 fare well due to their similar spectral properties - see Figs.~\ref{fig_comp_flux_max}\,$-$\,\ref{fig_comp_flux_max_he}). As discussed in Section~\ref{sect_spec}, these lines are not limited to He\one\,5876, 6678, and 7065\,\AA. We also identify He\one\ lines at 4471, 4922 and 5016. Hence, in what we could call ``genuine'' SNe Ib, most optical lines are due to He\one. Other lines are due to Ca\two\,H\&K, Fe\two\,5169\,\AA, O\one\,7774\,\AA\, and the Ca\two\ near-infrared triplet. The spectral properties are very similar for this sample of SNe Ib at early times post $V$-band maximum. One discrepancy is the much larger line width in SN\,2004gq compared to model he4 but this stems primarily from the very short rise time  of about 13\,d in SN\,2004gq, thus 10\,d less than in model he4. We also find that SN\,2005hg can be well fitted with model he6, with an offset of $-0.35$\,mag in $V$ (the SN is a little brighter than the model at the post-maximum epoch). The good match to the optical color suggests that our adopted total reddening of $E(B-V)=$\,0.09\,mag is suitable. The choice of $E(B-V)=$\,0.685\,mag by \citet{lyman_ibc_16} is an error and it is the origin of their inference of a large peak magnitude of $-19.2$\,mag and the erroneous status of prominent outlier for SN\,2005hg in the type Ib SN class. Instead, SN\,2005hg appears as a standard Type Ib SN.

For SNe\,2009jf and 2017ein, we used model he9 and he9pml -- progenitor model he9 is intermediate between Ib and Ic (see section~\ref{sect_prog}). Overall, the models reproduce most of the features although the model he9pml is too faint compared to SN\,2009jf. Matching the line widths is difficult since some are overestimated while others are underestimated by the model. This may be caused by the adopted mixing \citep{d12_snibc}, but may also be a signature of asphericity, for which we have unambiguous evidence in stripped-envelope SNe \citep{rest_casA_echo_11}.

For SNe\,2007gr and 2004aw, we use models that show no He\one\ lines (see Fig.~\ref{fig_comp_flux_max_he}). Model he10 fits well SN\,2007gr (despite the offset in rise time to maximum) while the comparison to SN\,2004aw required the higher energy model he12ep, in agreement with previous studies \citep{taubenberger_04aw_06,mazzali_04aw_17}. We also tried models with magnetar power. Model he12pml is a little too faint and underestimates line widths. Model he12pmu matches the brightness but is too blue and also underestimates line widths.

As far as we know, this is the first time that a single set of progenitor models, computed consistently from stellar evolution and exploded in a standard fashion (here using a thermal bomb to yield energies commensurate with theoretical expectations), reproduces the dichotomy between SNe Ib and SNe Ic spectral properties, as well as the photometric properties of a representative SN Ibc sample. The discrepancy in rise time between our models and some observations suggest lower ejecta masses, likely produced by binary mass exchange after He-core burning (as would be produced by case BB mass transfer).

\section{Discussion}
\label{sect_disc}

The simulations we present in this paper show a degeneracy in light curve and spectral properties for ejecta with a comparable $E_{\rm kin} / M_{\rm ej}$.  The representative ejecta expansion rate being the same for ejecta with the same $E_{\rm kin} / M_{\rm ej}$, the spectral line widths are also comparable. It is therefore challenging to separately constrain $E_{\rm kin}$  and $M_{\rm ej}$. There is no unique ejecta model to a given light curve and spectrum at maximum. Another way of stating this is that for any given model, one can find a counterpart at lower or higher $E_{\rm kin}$ (or $M_{\rm ej}$) that will exhibit the same light curve and the same spectrum at maximum, and potentially the same spectral evolution at all times.\footnote{A similar degeneracy was observed in the grid of SN Ic simulations presented in \citet{dessart_98bw_17} and \citet{D15_WR}. For example, one ejecta model with $E_{\rm kin}=4 \times 10^{51}$\,erg and $M_{\rm ej}=10$\,\msun\ yielded the same photometric and spectroscopic properties from 3 to 200\,d after explosion as a model counterpart with  $E_{\rm kin}=1.1 \times 10^{51}$\,erg and $M_{\rm ej}=5$\,\msun.} In practice, this may be complicated further by the presence of a chemical stratification and mixing, in particular because of the influence of \nifs.  The mixing of \nifs\ impacts the broadening of line profiles, even for the same ejecta expansion rate \citep{d12_snibc}, and can therefore impact our estimate of the energetics of the explosion.

Our modeling approach is to seek consistency by relying on physically-consistent progenitor models (which we argue here must arise from binary evolution for standard SNe Ibc) and reasonably motivated explosion physics (arguing for explosion energies in excess of $2 \times 10^{51}$\,erg may be challenging for standard SNe Ibc). The direct determination of the mass of various metals like O and Mg from nebular-phase spectra can also help to constrain the ejecta and progenitor masses, and therefore validate or invalidate what was inferred from the photospheric phase modeling (see, for example, \citet{jerkstrand_15_iib} for a study of SNe IIb).

Besides these conceptual difficulties, there are also uncertainties related to the light curve modeling itself. For example, we have shown in \citet{D15_SNIbc_I,D16_SNIbc_II} that the widely used Arnett model \citep{arnett_82} suffers from numerous limitations, which was confirmed and further documented by \citet{khatami_kasen_lc_19}. This impacts the inferred \nifs\ mass, as well as the ejecta mass, and therefore numerous other dependent quantities. Simplified light curve models fail to grasp potentially important physics (e.g., the role of mixing, the progenitor structure, the influence of non-LTE effects on the ionization etc) and thus retain uncertain systematic errors. One can generally tweak their model parameters one at a time (and thus independently) to obtain a perfect match to the light curve, but there is no guarantee that the successful parameters are adequate and consistent since they are integrated in a simplified model. In our approach with \cmfgen, a perfect match is generally impossible to obtain because of the large observational diversity and the substantial computational time taken by the simulations : we compute a handful of models to address the large diversity of observed SNe. Instead, we seek for a global consistency between our models and observations, and prefer to focus on trends obtained with grids of progenitor models obtained in a consistent fashion. It is the global consistency of our forward-modeling approach that reduces the otherwise vast parameter space.

\section{Conclusions}
\label{sect_conc}

We have presented a set of non-LTE time-dependent radiative transfer simulations for explosions arising from He star progenitors evolved with mass loss and at solar metallicity. By using such initial conditions,  we assume that the progenitor H-rich envelope was promptly and entirely removed by binary mass exchange at core He ignition. Using a new prescription for WR star wind mass loss \citep{yoon_wr_17}, these He star models reach core collapse with a total mass in the range 3.0 to 5.4\,\msun, hence a tighter range than obtained for the same range in initial mass with alternate mass loss prescriptions or if the star had evolved in isolation from the zero-age main sequence. Besides offering a means to produce the faint end of the distribution of WC star luminosities \citep{sander_wc_12}, it also opens up the possibility of producing SN Ic ejecta masses more in line with those inferred from observations.

In a first set of explosion models in which we impose an ejecta kinetic energy of $5 \times 10^{50}$\,erg\,$M_\odot^{-1}$ and the same \nifs\ mass of 0.08\,\msun, the resulting light curves of He-rich (he4\,$-$\,he8) and He-poor models (he10\,$-$\,he12) are nearly identical. This ansatz requires that the explosion mechanism delivers about two to three times more energy in higher mass progenitors than in lower mass ones, which is in line with theoretical predictions \citep{ertl_ibc_20}. These simulations reach a peak luminosity of about 10$^{42.2}$\,\ergs, which is representative of the majority of SNe Ibc.

Our models systematically peak at about 20\,d after explosion. While it is customary in the community to invoke very energetic explosions to explain even garden-variety SNe Ibc, we suspect that the short rise time of some SNe Ibc arises from progenitors that suffered two major events of binary mass exchange, when the star first became a supergiant (case B mass transfer) and later when the H-deficient star expanded during the strong core contraction at the end of its life (case BB mass transfer). Hence, these events indicate that SN Ibc ejecta masses can be very low (by massive star standards), probably reaching down to $\lesssim$\,1\,\msun\ in all the fast rising SNe Ibc. The small number of SNe Ibc with a short rise time appears consistent with the small number of binaries expected to undergo Case BB mass transfer.

To cover a broader range of peak luminosities, some models were exploded with a greater energy, or were assigned a greater \nifs\ mass, or finally were influenced by power from the compact remnant (treated in this study as magnetar power, but it could equivalently be associated with power from fallback accretion).  Doubling $E_{\rm kin}$, doubling the \nifs\ mass, or introducing a modest power from the compact remnant can extend the peak luminosities of our models to 10$^{42.6}$\,\ergs. These ingredients impact the late-time light curve differently. A higher expansion rate favors $\gamma$-ray escape so that \nifs-powered ejecta tend to show a faster decline rate at nebular times in this case. In contrast, ejecta influenced by a magnetar reach higher luminosities both at peak and at nebular times, exhibiting a slower decline at late times because of the full trapping of magnetar power. Overall, the combination of these parameters ($E_{\rm kin}$, \nifs\ mass, power injection from the compact remnant) yields a wide diversity of light curve properties that overlaps with the observed parameter space occupied by the majority of SNe Ib and Ic.

We have also revised downward the inferred luminosities of a number of SNe Ibc, as a result of a reduction in the inferred reddening. For example, for SN\,2005hg, \citet{lyman_ibc_16} propose a huge \nifs\ mass of 0.66\,\msun\ and identify that SN Ib as an outlier. This arises from their inferred reddening $E(B-V)$ of 0.685\,mag. With our models, we match the color of SN\,2005hg with $E(B-V)$ of 0.1\,mag, which places SN\,2005hg as a standard SN Ib. Similarly, the large reddenings used in the literature for SNe 2007C and 2007uy are probably overestimated (see section~\ref{sect_spec}). Finally, numerous SNe Ib (e.g. SN\,2005bf) or Ic (e.g., PTF12gzk) exhibit clear peculiarities that suggest they differ in a fundamental way from the majority of SNe Ibc (the same way as interacting type II SNe cannot be analyzed and interpreted as part of a Type II-P SN sample). Failing to treat these apart, one tends to extend the distribution of SN Ibc peak luminosities to large values (and consequently their inferred \nifs\ mass since only \nifs\ is invoked to explain their luminosity), the same way the distribution of SN II luminosities would extend to large values if one were to group SNe II-P with the SNe IIn and treat them as one sample.

Spectroscopically, the dichotomy between Type Ib and Type Ic SNe is reproduced in our model sequence he4 to he12, all characterized by a strong \nifs\ mixing. Progenitors that have retained their He-rich shell (where He has a 98\% mass fraction; these progenitors correspond to WN stars) show He\one\ lines around bolometric maximum (they would show He\one\ lines throughout the photospheric phase; \citealt{dessart_11_wr}). Progenitors that have lost this He-rich shell still contain some He (at the $\sim$\,10\% level), but mixed with C and O, and produce optical spectra with no strong He\one\ line. In these, He\one\,5876 and 7065\,\AA\ are predicted but their strength is very weak (i.e., 1\% above the continuum). This exploration would suggest that technically, all these models are SNe Ib, but this classification is then useless to understand the properties and the origin of SNe Ib since they would then encompass all possible WR star progenitors. A more useful classification for SN science would be to define Type Ib SNe as events in which many He\one\ lines are present in the optical, and at least 5876, 6678, and 7065\,\AA. In particular, SNe showing no He\one\,6678\,\AA\ are most likely very deficient in He.\footnote{He\one\,6678\,\AA\ arises from more excited, and therefore less populated, states. It is consequently weaker than He\one\,5876\,\AA. In our simulations, we find that this line is only present at bolometric maximum if the progenitor star has a He-rich shell, i.e., wherein the He mass fraction is $1-Z$.} Hence, our stellar evolution, stellar explosion, and radiative-transfer simulations suggest that one may explain SNe Ib as arising from He-star models that have retained their He-rich shell at core collapse, while SNe Ic arise from the more massive He-star models that reach core collapse with little He left.

While binarity appears as an essential channel for the formation of these He stars, observed SNe Ibc probably suggest additional bifurcations for the progenitor evolution. At the low mass end, low mass He giants may form and produce some SNe Ib \citep{bersten_iPTF13bvn_14,eldridge_13bvn_15,eldridge_maund_16,d18_ext_ccsn}. Some low mass He stars have been observed \citep{neugent_wn3o3_17, smith_wn3o3_18}. Combined with a tight orbit, this late expansion of the star may lead to  the mass transfer of the He-rich shell (case BB) and produce an ultra-stripped SN Ic \citep{dewi_bin_02,dewi_pols_03,tauris_ulstr_13}. At higher He-star mass, case BB mass transfer could produce events like SNe\,1994I or 2007gr in which the ejecta mass is around 1\,\msun\ \citep{nomoto_94I_94,tauris_ulstr_15}. Case BB mass transfer is, however, rare and cannot be at the origin of the majority of SNe Ic.

As discussed above, the single star channel, as well as various binary channels,  may contribute to produce SNe\,Ibc. However, binary evolution predicts that the majority of SNe\,Ibc stem from stable Case\,A or Case\,B mass transfer, which is inevitable for roughly half of the massive binaries (with most of the other half merging to single stars; \citealt{demink_merger_14}; \citealt{wang_bin_ms_20}). This allows for a simple prediction of the metallicity dependence of the number of SNe Ibc. Since binary stripping does not, to first order, depend on metallicity, our results would not predict a $Z$-dependence of the overall Ibc rate. However, after the binary stripping, the loss of the helium layer is mediated by $Z$-dependent winds. Consequently, we would expect an increase of the Ic/Ib SN number ratio with metallicity (see, for example, \citealt{georgy_snibc_09}). The findings of \citet{anderson_env_ccsn_15}, who study the location of different types of SNe in morphologically normal and disturbed host galaxies, appear to be consistent with this expectation. Firmer conclusions require a detailed investigation of models with various metallicities, which we will perform in the near future.

An aspect not discussed in the present study, but already discussed in previous studies with \cmfgen, is the difficulty of reproducing the early time properties of SNe Ibc. In general, bright and blue SNe IIb and Ib are easily explained by invoking the explosion of supergiant or giant stars (with some residual H or not), which are both predicted by binary star evolution for low to moderate mass He stars. However, numerous SNe Ib and Ic exhibit blue optical colors at early times without an appreciable luminosity boost. In the sample of observations selected for this paper, most of our He star models are redder (and sometimes fainter) than observed. Our spectra also show broader lines than usually observed, indicating their lack of material beyond a certain velocity. This suggests that something is missing in our WR star models. Possible solutions are wave excitation (causing inflation, partial unbinding of the envelope, or a strong wind; see for example \citealt{fuller_ro_18}) or nuclear flashes \citep{woosley_he_19}.   The offset in color and brightness would then be resolved by invoking a more extended progenitor star \citep{nomoto_93j_93, podsiadlowski_93j_93,blinnikov_94_93j,woosley_94_93j} or interaction with circumstellar material, as evidenced in an extreme way in SNe Ibn like 2006jc \citep{pasto_06jc_07}. Other complications such as mixing and asymmetry may likely influence standard SNe Ibc although it is not clear how much diversity these effects could produce (mixing is probably strong in all SNe Ibc and asymmetry should be significant only in the most extreme explosions leading to GRB/SNe).

Our study is the first to show that non-LTE time-dependent radiative transfer models based on He-star explosions (themselves based on a set of physically consistent stellar evolution and explosion models) can reproduce the photometric properties of the majority of SNe Ibc, as well as the observed dichotomy between the spectral properties of Type Ib and Ic SNe. These models indicate a compatibility between stellar evolution, stellar explosion, and SN Ib and Ic observations, thereby resolving a long standing issue.

\begin{acknowledgements}

SCY is supported by the National Research Foundation of Korea (NRF) grant (NRF- 2019R1A2C2010885).
This work was granted access to the HPC resources of  CINES under the allocations  2018 -- A0050410554 and 2019 -- A0070410554 made by GENCI.

\end{acknowledgements}

\end{document}

%% file: table_prop_mesa_heprog.tex
\begin{table*}
\caption{Properties of our set of progenitor models. Models he4-he8 (he10-he12) correspond to SNe Ib (Ic) progenitors. Model he9 is an intermediate case.
\label{tab_prog}
}
\begin{center}
\begin{tabular}{
l@{\hspace{4mm}}c@{\hspace{4mm}}c@{\hspace{4mm}}
c@{\hspace{4mm}}c@{\hspace{4mm}}c@{\hspace{4mm}}
c@{\hspace{4mm}}c@{\hspace{4mm}}c@{\hspace{4mm}}
c@{\hspace{4mm}}c@{\hspace{4mm}}c@{\hspace{4mm}}
c@{\hspace{4mm}}c@{\hspace{4mm}}c@{\hspace{4mm}}
}
\hline

Model & $M_{\rm i}$ & $M_{\rm f}$ & $M_{\rm Fe,c}$ & $M_{\rm CO,c}$ &  $\log L_\star$ & $R_\star$ &    $M$(He) &  $M$(C) & $M$(N) & $M$(O) &  $X_s$(He) &   $X_s$(C) &   $X_s$(N) &   $X_s$(O) \\
      &   [\msun]  &   [\msun]  & [\msun]  &[\msun]  & [\lsun]      &  [\rsun] &     [\msun]  &[\msun]  & [\msun]  &[\msun]  & & & & \\
\hline
  he4 &  4.0   &   2.98 &  1.46     &      1.82   &    4.57 &     7.97 &   1.133 &      0.056 &      0.009 &      0.150 &      0.980 &      0.000 &      0.013 &      0.000 \\
  he5 &  5.0   &   3.56 &  1.47     &      2.29   &    4.79 &     3.60 &   1.195 &      0.110 &      0.008 &      0.423 &      0.980 &      0.000 &      0.013 &      0.000 \\
  he6 &  6.0   &   4.07 &  1.80     &      2.71   &    4.92 &     2.46 &   1.160 &      0.227 &      0.006 &      0.571 &      0.980 &      0.000 &      0.013 &      0.000 \\
  he7 &  7.0   &   4.55 &  1.63     &      3.13   &    5.01 &     1.51 &   1.075 &      0.352 &      0.005 &      0.838 &      0.980 &      0.000 &      0.013 &      0.000 \\
  he8 &  8.0   &   5.00 &  1.64     &      3.55   &    5.09 &     1.23 &   0.935 &      0.503 &      0.003 &      1.214 &      0.980 &      0.000 &      0.013 &      0.000 \\
\hline                                                                     
 he9 &  9.0    &   5.29 &   1.57    &       3.93  &    5.12 &     0.87 &   0.628 &      0.659 &      0.000 &      1.487 &      0.822 &      0.148 &      0.000 &      0.005 \\
\hline                                                                     
 he10 & 10.0   &   5.20 &  1.67     &      4.06   &    5.16 &     0.46 &   0.327 &      0.678 &      0.000 &      1.580 &      0.452 &      0.452 &      0.000 &      0.071 \\
 he11 & 11.0   &   5.25 &  1.55     &      4.01   &    5.12 &     0.59 &   0.220 &      0.674 &      0.000 &      1.659 &      0.295 &      0.525 &      0.000 &      0.155 \\
 he12 & 12.0   &   5.43 &  1.67     &      4.11   &    5.20 &     0.49 &   0.236 &      0.763 &      0.000 &      1.831 &      0.195 &      0.530 &      0.000 &      0.251 \\
\hline
\end{tabular}
\end{center}
{\bf Notes:} For each progenitor model, the table gives the initial mass $M_{\rm i}$, the final mass $M_{\rm f}$, the Fe-core mass (where the electron fraction first rises to 0.499 from the center), the CO-core mass (outermost location where the He mass fraction is lower than 0.01), the final surface luminosity $\log L_\star$, the final surface radius $R_\star$, followed by the total mass of He, C, N, O, as well as the surface mass fraction of He, C, N, and O.
\end{table*}

%% file: table_prop_ejecta.tex
\begin{table}
\caption{Properties of ejecta models produced from the set of progenitor models presented in Section~\ref{sect_prog} and Table~\ref{tab_prog}.
\label{tab_ejecta}
}
\begin{center}
\begin{tabular}{
l|cc@{\hspace{1mm}}
c@{\hspace{1mm}}c@{\hspace{1mm}}c@{\hspace{1mm}}
c@{\hspace{1mm}}c@{\hspace{1mm}}
}
\hline
       Model  &   $M_{\rm ej}$ &  $E_{\rm kin}$     & $V_m$   &  $^{56}{\rm Ni}_0$  &  $E_{\rm pm}$   \\
\hline
              &  [\msun]       &  [10$^{51}$\,erg]  &  [\kms]   &      [\msun]  &  [10$^{49}$\,erg]    \\
\hline
      he4       &    1.49    &       0.75   &    7117     &      0.08     &  $\cdots$   \\
      he5       &    2.01    &       1.02   &    7134     &      0.08     &  $\cdots$  \\
      he6       &    2.37    &       1.21   &    7152     &      0.08     &  $\cdots$  \\
      he7       &    2.85    &       1.46   &    7187     &      0.08     &  $\cdots$  \\
      he8       &    3.31    &       1.67   &    7135     &      0.08     &  $\cdots$  \\
      he9       &    3.59    &       1.81   &    7119     &      0.08     &  $\cdots$  \\
     he10       &    3.39    &       1.77   &    7239     &      0.08     &  $\cdots$  \\
     he11       &    3.56    &       1.77   &    7068     &      0.08     &  $\cdots$  \\
     he12       &    3.67    &       2.11   &    7608     &      0.08     &  $\cdots$  \\
\hline
      he4       &    1.49    &       0.75   &    7117     &      0.08     &  $\cdots$   \\
   he4epl       &    1.49    &       1.63   &   10504     &      0.08     &  $\cdots$  \\
    he4ep       &    1.49    &       1.64   &   10509     &      0.16     &  $\cdots$  \\
\hline
   he4pml       &    1.49    &       0.75   &    7118     &      0.08     &  1.0       \\
   he4pmu       &    1.49    &       0.75   &    7124     &      0.08     &  2.0       \\
\hline
      he9       &    3.59    &       1.81   &    7119     &      0.08     &  $\cdots$  \\
   he9pml       &    3.59    &       1.81   &    7121     &      0.08     &  1.0       \\
   he9pmu       &    3.59    &       1.81   &    7123     &      0.08     &  2.0       \\
\hline
     he12       &    3.67    &       2.11   &    7608     &      0.08     &  $\cdots$  \\
  he12epl       &    3.68    &       4.77   &   11414     &      0.08     &  $\cdots$  \\
   he12ep       &    3.69    &       4.76   &   11396     &      0.16     &  $\cdots$  \\
\hline
  he12pml       &    3.67    &       2.11   &    7604     &      0.08     &  1.0       \\
  he12pmu       &    3.67    &       2.11   &    7605     &      0.08     &  2.0       \\
he12eppml       &    3.68    &       4.76   &   11409     &      0.16     &  1.0       \\
\hline
\end{tabular}
\end{center}
{\bf Notes:} For models with a suffix pml and pmu, the magnetar field is 10$^{15}$\,G.
\end{table}

%% file: table_prop_sn_radiation.tex
\begin{table*}
\caption{Properties of the $UVOIR$ and $V$-band light curves for the full model set, including the rise time to maximum, the peak luminosity or magnitude, and the
decline in brightness between the time of maximum and 15 days later. The last five columns give the temperature, velocity, overlying mass as well as the He and O
mass fractions at the photosphere (defined on a Rosseland-mean optical depth scale) at the time of bolometric maximum.
\label{tab_sn_rad}
}
\begin{center}
\begin{tabular}{
l|c@{\hspace{4mm}}c@{\hspace{4mm}}c@{\hspace{4mm}}|
c@{\hspace{4mm}}c@{\hspace{4mm}}c@{\hspace{4mm}}|
c@{\hspace{4mm}}c@{\hspace{4mm}}c@{\hspace{4mm}}
c@{\hspace{4mm}}c@{\hspace{4mm}}
}
\hline
       Model  &  \multicolumn{3}{c|}{$L_{UVOIR}$} & \multicolumn{3}{c|}{V} &            $T_{\rm ph}$ & $V_{\rm ph}$ & $\Delta M_{V > V_{\rm ph}}$ & $X$(He)$_{\rm ph}$ & $X$(O)$_{\rm ph}$ \\
\hline
              &   $t_{\rm rise}$  &  Max &  $\Delta M_{15}$  &  $t_{\rm rise}$  &  Max &  $\Delta M_{15}$  &      & & &  & \\
              &     [d]           & [10$^{42}$\,erg\,s$^{-1}$]&  [mag]    & [d]  & [mag]   & [mag]          & [K] & [\kms] & [\msun] & & \\
\hline
      he4       &       23.0    &      1.60    &      0.68   &      22.4   &     -17.12   &       0.82  &     6448    &   8394  &   0.36 &   0.94  &    0.00     \\
      he5       &       22.2    &      1.68    &      0.69   &      21.4   &     -17.17   &       0.84  &     6355    &   8896  &   0.39 &   0.93  &    0.00    \\
      he6       &       22.4    &      1.61    &      0.63   &      21.7   &     -17.15   &       0.82  &     6287    &   8819  &   0.48 &   0.85  &    0.02    \\
      he7       &       22.2    &      1.60    &      0.59   &      21.1   &     -17.15   &       0.70  &     6335    &   8637  &   0.60 &   0.70  &    0.06    \\
      he8       &       22.5    &      1.55    &      0.50   &      21.1   &     -17.10   &       0.59  &     6220    &   8617  &   0.65 &   0.57  &    0.07    \\
      he9       &       23.7    &      1.46    &      0.42   &      21.7   &     -17.06   &       0.50  &     6035    &   8297  &   0.80 &   0.33  &    0.17    \\
     he10       &       23.5    &      1.49    &      0.50   &      21.6   &     -17.12   &       0.59  &     6022    &   8427  &   0.76 &   0.18  &    0.28    \\
     he11       &       24.4    &      1.48    &      0.50   &      22.2   &     -17.09   &       0.59  &     6000    &   8344  &   0.76 &   0.11  &    0.39    \\
     he12       &       23.5    &      1.48    &      0.50   &      21.5   &     -17.12   &       0.58  &     5920    &   8758  &   0.86 &   0.11  &    0.37    \\
\hline
      he4       &       23.0    &      1.60    &      0.68   &      22.4   &     -17.12   &       0.82  &     6448    &   8394  &   0.36 &   0.94  &    0.00     \\
   he4epl       &       18.2    &      1.76    &      0.87   &      17.8   &     -17.20   &       1.02  &     6581    &  12351  &   0.36 &   0.94  &    0.00    \\
    he4ep       &       20.3    &      3.07    &      0.69   &      20.3   &     -17.80   &       0.82  &     7240    &  12805  &   0.32 &   0.93  &    0.00    \\
\hline
      he4       &       23.0    &      1.60    &      0.68   &      22.4   &     -17.12   &       0.82  &     6448    &   8394  &   0.36 &   0.94  &    0.00     \\
   he4pml       &       25.8    &      2.19    &      0.44   &      25.4   &     -17.44   &       0.55  &     6717    &   8208  &   0.38 &   0.94  &    0.00    \\
   he4pmu       &       27.9    &      3.38    &      0.40   &      27.6   &     -17.83   &       0.42  &     7556    &   8297  &   0.37 &   0.94  &    0.00    \\
\hline
      he9       &       23.7    &      1.46    &      0.42   &      21.7   &     -17.06   &       0.50  &     6035    &   8297  &   0.80 &   0.33  &    0.17    \\
   he9pml       &       26.5    &      2.15    &      0.33   &      24.5   &     -17.50   &       0.41  &     6420    &   8108  &   0.86 &   0.33  &    0.18    \\
   he9pmu       &       27.7    &      3.48    &      0.36   &      27.1   &     -18.01   &       0.42  &     7373    &   8471  &   0.76 &   0.34  &    0.17    \\
\hline
     he12       &       23.5    &      1.48    &      0.50   &      21.5   &     -17.12   &       0.58  &     5920    &   8758  &   0.86 &   0.11  &    0.37    \\
  he12epl       &       18.2    &      1.78    &      0.76   &      16.5   &     -17.33   &       0.87  &     5877    &  12617  &   0.95 &   0.09  &    0.32    \\
   he12ep       &       19.8    &      3.26    &      0.62   &      18.1   &     -18.04   &       0.72  &     6489    &  12876  &   0.90 &   0.09  &    0.33    \\
\hline
     he12       &       23.5    &      1.48    &      0.50   &      21.5   &     -17.12   &       0.58  &     5920    &   8758  &   0.86 &   0.11  &    0.37    \\
  he12pml       &       26.5    &      2.16    &      0.36   &      24.4   &     -17.55   &       0.44  &     6235    &   8583  &   0.91 &   0.10  &    0.37    \\
  he12pmu       &       28.4    &      3.43    &      0.36   &      26.9   &     -18.07   &       0.44  &     6908    &   8573  &   0.92 &   0.10  &    0.37    \\
he12eppml       &       21.9    &      3.75    &      0.49   &      20.2   &     -18.19   &       0.57  &     6573    &  12373  &   1.00 &   0.09  &    0.31    \\
\hline
\end{tabular}
\end{center}
\end{table*}